\documentclass[journal=cmatex,manuscript=article]{achemso}

\usepackage[T1]{fontenc}
\usepackage[utf8]{inputenc}
\usepackage[table]{xcolor}
\usepackage{chemformula}
\usepackage{xr}
\usepackage[pdfborder={0 0 0}, pdfencoding=auto, pdfa]{hyperref}
\usepackage{cleveref}
\usepackage{siunitx}
\usepackage{xspace}
\usepackage{multirow}
\usepackage{hhline}
\usepackage{csquotes}
\usepackage{comment}
\usepackage{amsmath,bbm,bm,amssymb}
\usepackage{longtable}
\usepackage{lmodern}

\DeclareSIUnit\at{atom}
\DeclareSIUnit\fu{f.u.}
\DeclareSIUnit\atm{atm}

\AtBeginDocument{\singlespacing}

\newcolumntype{L}[1]{>{\raggedright\arraybackslash}p{#1}} 	
\newcolumntype{C}[1]{>{\centering\arraybackslash}p{#1}} 		
\newcolumntype{R}[1]{>{\raggedleft\arraybackslash}p{#1}} 		
\newcolumntype{P}[1]{>{\hspace{0pt}}p{#1}}

\makeatletter
\renewcommand*{\acs@author@fnsymbol@symbol}[1]{%
  \ifnum#1=0 
    *%
  \else%
    [\@alph{#1}]%
  \fi%
}
\makeatother

\author{Pascal Henkel}
\affiliation{Department of Applied Physics, Aalto University, P.O.Box 11100, FI-00076 AALTO, Finland}\email{pascal.henkel@aalto.fi}

\author{G. Krishnamurthy Grandhi}
\affiliation{Hybrid Solar Cells, Faculty of Engineering and Natural Sciences, Tampere University, FI-33014 Tampere, Finland}

\author{Jarno Laakso}
\affiliation{Department of Applied Physics, Aalto University, P.O.Box 11100, FI-00076 AALTO, Finland}

\author{David Rovira Ferrer}
\affiliation{Micro and Nanotechnology Group, Emerging Thin Film Photovoltaics Lab, Universitat Polit\`{e}cnica de Catalunya (UPC), ES-08019 Barcelona, Spain; Barcelona Research Center in Multiscale Science and Engineering, Universitat Polit\`{e}cnica de Catalunya (UPC), ES-08019 Barcelona, Spain}

\author{Edgardo Saucedo}
\affiliation{Micro and Nanotechnology Group, Emerging Thin Film Photovoltaics Lab, Universitat Polit\`{e}cnica de Catalunya (UPC), ES-08019 Barcelona, Spain; Barcelona Research Center in Multiscale Science and Engineering, Universitat Polit\`{e}cnica de Catalunya (UPC), ES-08019 Barcelona, Spain}

\author{Jingrui Li}
\affiliation{State Key Laboratory for Manufacturing Systems Engineering; Electronic Materials Research Laboratory, Key Laboratory of the Ministry of Education, School of Electronic Science and Engineering; International Joint Laboratory for Micro/Nano Manufacturing and Measurement Technology, Xi'an Jiaotong University, Xi'an CN-710049, China}

\author{Paola Vivo}
\affiliation{Hybrid Solar Cells, Faculty of Engineering and Natural Sciences, Tampere University, FI-33014 Tampere, Finland}

\author{Miguel A. L. Marques}
\affiliation{Research Center Future Energy Materials and Systems of the University Alliance Ruhr and Interdisciplinary Centre for Advanced Materials Simulation, Ruhr University Bochum, DE-44801 Bochum, Germany}

\author{Patrick Rinke}
\affiliation{Department of Applied Physics, Aalto University, P.O.Box 11100, FI-00076 AALTO, Finland}
\alsoaffiliation{Atomistic Modelling Center, Munich Data Science Institute, Technical University of Munich, DE-85748 Garching bei M\"unchen, Germany; Munich Center for Machine Learning (MCML), DE-80333 M\"unchen, Germany; Physics Department, TUM School of Natural Sciences, Technical University of Munich, DE-85748 Garching bei M\"unchen, Germany}
\email{patrick.rinke@tum.de}

\title{How Thermodynamically Accessible are Quaternary Mixed-Metal Chalcohalides?}

\abbreviations{mixed-metal chalcohalides, thermodynamic accessibility, compositional engineering, density functional theroy, molecular ink synthesis}

\begin{document}
\begin{abstract}
Quaternary mixed-metal chalcohalides (MMCHs) are a promising mixed-anion semiconductor family, but their thermodynamic accessibility remains largely unexplored. We combine density functional theory with the Alexandria Materials Database to carry out a rapid and reliable evaluation of the thermodynamic accessibility of 54 \ch{M(II)2M(III)Ch2X3} compounds. All evaluated MMCHs are predicted to be thermodynamically unstable, but a considerable fraction of MMCHs lie within computational uncertainty of the convex hull. Their predicted decomposition follows five distinct reaction pathways, most commonly into \ch{M(III)2Ch3}, \ch{M(II)X2}, and \ch{M(II)Ch}. Solution processing of six Sn-based MMCHs supports our analysis. Notably, this includes the first synthesis of the previously unreported \ch{Sn2InS2Br3} in a multiphase film. Our approach identifies convex hull proximity and competing phase formation as key constraints on MMCH synthetic accessibility during solution processing, thereby guiding future MMCH exploration.\end{abstract}


\section{1 Introduction}
Photovoltaic (PV) technologies are a cornerstone of sustainable energy production, providing a scalable source of clean, affordable, and secure energy. Their widespread deployment requires absorber materials that combine high power conversion efficiencies (PCEs), low manufacturing costs, and long-term stability. In this context, metal halide perovskites have attracted significant attention as promising candidates.\cite{feng2023,zhang2016,NRELchart} In particular, lead halide perovskites (LHPs) combine cost-effective fabrication with favourable optoelectronic properties, including a high degree of defect tolerance, and have demonstrated astonishing PV performance in laboratory conditions.\cite{seo2016,nie2020a,savill2021} However, their practical application remains limited by lead toxicity and moderate long-term stability under operating conditions.\cite{huang2021, huang2021a, ganose2017, nie2020a} Low-toxicity alternatives based on \ch{Sn^{2+}}, \ch{Sb^{3+}}, and \ch{Bi^{3+}} have therefore been extensively investigated, but often suffer from intrinsic limitations such as high defect densities, ultrafast carrier localisation, and, for \ch{Sn}-based compounds, rapid oxidation of \ch{Sn^{2+}} to \ch{Sn^{4+}} under ambient conditions.\cite{cao2021, grandhi2024} Metal chalcogenides based on \ch{Pb^{2+}}, \ch{Cd^{2+}}, or \ch{Sb^{3+}} offer a complementary route, combining strong absorption, tunable band gaps, and high chemical stability, although their PCEs remain modest.\cite{NRELchart,choi2014,im2011,yadav2023} This motivates the search for absorbers that combine the favourable electronic and processing properties of halides with the chemical robustness of chalcogenides.

Quaternary mixed-metal chalcohalides (MMCHs) have emerged as promising PV absorbers, and, more broadly, as a chemically versatile mixed-anion platform in which metal-halide and metal-chalcogen coordination environments found in halide perovskites (\ch{ABX3}) and metal chalcogenides (\ch{MCh2}) coexist within a single crystalline phase.\cite{nie2020b, kavanagh2021b, sopiha2022}. Their stoichiometry is \ch{M(II)2M(III)Ch2X3}, where the \ch{M(II)} and \ch{M(III)} sites are typically occupied by $ns^2$ lone-pair metals, the \ch{Ch} site by chalcogens, and the \ch{X} site by halogens. Recently, a \ch{Sn2SbS2I3}-based device has achieved a PCE of \SI{4.04}{\percent},\cite{nie2020b} which, although still well below state-of-the-art LHP solar cells (which now exceed \SI{26}{\percent}\cite{NRELchart}), represents a promising starting point.

MMCHs can crystallise in three different space groups, namely $Cmcm$,\cite{olivierfourcade1980, ibanez1984, islam2016, dolgikh1985, doussier2007, roth2023} $Cmc2_1$,\cite{kavanagh2021b} and $P2_1/c$.\cite{doussier2007} The properties of the perovskite and chalcogenide building blocks are expected to combine synergistically in MMCHs, giving rise to high defect tolerance through strong dielectric screening from $ns^2$ lone-pair electrons, low defect capture cross sections, and dispersive valence and conduction band edges.\cite{huang2021, huang2021a, nie2020b} In addition, the strong metal-chalcogen bonds formed between the bivalent chalcogenide anions and both \ch{M(II)} and \ch{M(III)} cations are believed to overcome the stability limitations commonly observed in LHPs. The \ch{Sn-Ch} bonding motif, in particular, is expected to suppress oxidation from \ch{Sn^{2+}} to \ch{Sn^{4+}} in lead-free MMCHs. Consistently, \ch{Sn2SbS2I3} thin films synthesised under reducing conditions have been shown to remain stable even in humid environments, as verified by X-ray photoelectron spectroscopy.\cite{nie2020b, kavanagh2021b} Despite these promising results for \ch{Sn2SbS2I3}, systematic investigation of quaternary MMCH compounds remains scarce. To date, only five MMCH compounds have been experimentally synthesised: \ch{Sn2SbS2I3},\cite{olivierfourcade1980, nie2020b, kavanagh2021b, nicolson2023} \ch{Sn2SbSe2I3},\cite{ibanez1984} \ch{Sn2BiS2I3},\cite{islam2016} \ch{Pb2SbS2I3}\cite{dolgikh1985, doussier2007, roth2023} and \ch{Pb2BiS2I3}.\cite{islam2016, roth2023} This leaves an open question: How thermodynamically accessible are MMCHs?

Evaluating the stability of chemically complex materials, such as quaternary MMCHs, remains challenging both experimentally and theoretically. While recent advances in automated and high-throughput experimental platforms increasingly enable systematic experimental screening,\cite{angelopoulos2024, chen2026, zhang2024} theoretical approaches remain essential for efficiently narrowing the chemical search space.\cite{henkel2023, laakso2022, korbel2016} In particular, high-throughput computational assessments of stability against competing phases can provide valuable guidance by identifying promising compositions prior to experimental synthesis. A central challenge in the computational evaluation of material stability lies in the need for sufficiently large datasets required to accurately capture all relevant competing phases. Recent developments in computational materials databases, such as the Materials Project,\cite{materialsproject2013} AFLOW,\cite{curtarolo2012} NOMAD,\cite{draxl2018} and Alexandria,\cite{cavignac2026} enable the thermodynamic stability screening across much broader materials spaces.\cite{schmidt2022a, clary2020, bartel2019}

The aim of this study is to determine the thermodynamic accessibility of quaternary MMCHs and to identify the competing phases that limit their synthesis. We therefore investigate a chemically diverse set of 54 MMCH compounds comprising \ch{Sn^{2+}}- and \ch{Pb^{2+}}-based systems at the \ch{M(II)} site and \ch{Sb^{3+}}, \ch{Bi^{3+}}, or \ch{In^{3+}} at the \ch{M(III)} site, where all metals except \ch{In^{3+}} (with a $5d^{10}6s^0$ valence configuration) feature $ns^2$ lone-pair electrons. The \ch{Ch} site is occupied by \ch{S^{2-}}, \ch{Se^{2-}}, or \ch{Te^{2-}}, and the \ch{X} site by \ch{Cl^-}, \ch{Br^-}, or \ch{I^-}. We evaluate the thermodynamic accessibility of these compounds by combining computational and experimental approaches. Using density functional theory (DFT) in conjunction with the Alexandria Materials Database,\cite{cavignac2026} which contains \num{5.8} million DFT-optimised crystal structures and their energies, we evaluate the stability of MMCH compounds at \SI{0}{\kelvin} with respect to all competing phases. By leveraging the precomputed phases available in Alexandria, only a limited number of additional DFT calculations is required. Our approach facilitates a rapid yet chemically comprehensive convex hull analysis, the identification of decomposition products, and the systematic analysis of competing decomposition pathways. Combining all aspects, we achieve a quantitative assessment of the thermodynamic accessibility of MMCH compounds.

Building on these analyses and our previous MMCH studies,\cite{henkel2023, henkel2025} this work provides a systematic overview of convex hull energies and decomposition trends across the MMCH materials space, offering guidance for the targeted experimental synthesis of selected compounds. In particular, we distinguish whether an MMCH compound is a thermodynamic ground state (at \SI{0}{\kelvin}), whether it lies sufficiently close to the convex hull to remain experimentally plausible, and whether a given processing route can access and retain that phase before competing binary or ternary compounds form.

\section{2 Results and Discussion}
To examine the thermodynamic accessibility of MMCH compounds, we first computed their stability with respect to competing phases at \SI{0}{\kelvin}. To this end, we examined the chemical space spanned by the four constituent elements of the MMCH compound and determined the stability of all 54 compounds (see SI Section S1 for more details) in the three space groups $Cmcm$, $Cmc2_1$, and $P2_1/c$. This approach allows us to assess whether an MMCH compound is stable or prone to decomposition into competing phases, providing us with a robust overview of the thermodynamic accessibility across the MMCH chemical space. Subsequent analysis of the computed energies was conducted using random-forest (RF) regression models and Shapley additive explanations (SHAP) analysis to quantify the effect of various substitutional elements. Lastly, based on our theoretical insights, we selected six MMCH compounds (\ch{Sn2SbS2I3}, \ch{Sn2InS2Br3}, \ch{Sn2InSe2Br3}, \ch{Sn2SbS2Cl3}, \ch{Sn2SbSe2Cl3}, and \ch{Sn2InSe2Cl3}) for additional experimental investigation (see SI Section S2 for more details), which enabled us to study their experimental accessibility and stability.


\subsection{2.1 Compositional Phase Diagrams}
We determined the stability of MMCH compounds with respect to all \enquote{known} competing phases within the chemical space spanned by the four constituent elements (e.g., \ch{Sn}, \ch{In}, \ch{S}, and \ch{Br} for \ch{Sn2InS2Br3}). For all phases, the formation energies were calculated (see SI Equation\,(1) for details) with energies of the competing phases taken from the Alexandria Materials Database and those of the MMCH compounds obtained from our own DFT calculations. Based on these formation energies, we constructed the compositional phase diagrams and their associated convex hull at $T = \SI{0}{\kelvin}$ and $P = \SI{0}{\atm}$. The convex hull represents the set of energetically most stable phases at a given composition, thereby defining the ground states. Compounds located on the convex hull are thus thermodynamically stable, whereas those above the hull are unstable and prone to decomposition into a combination of stable competing phases.

\begin{figure}[H]
\begin{center}
\includegraphics[width=0.5\textwidth]{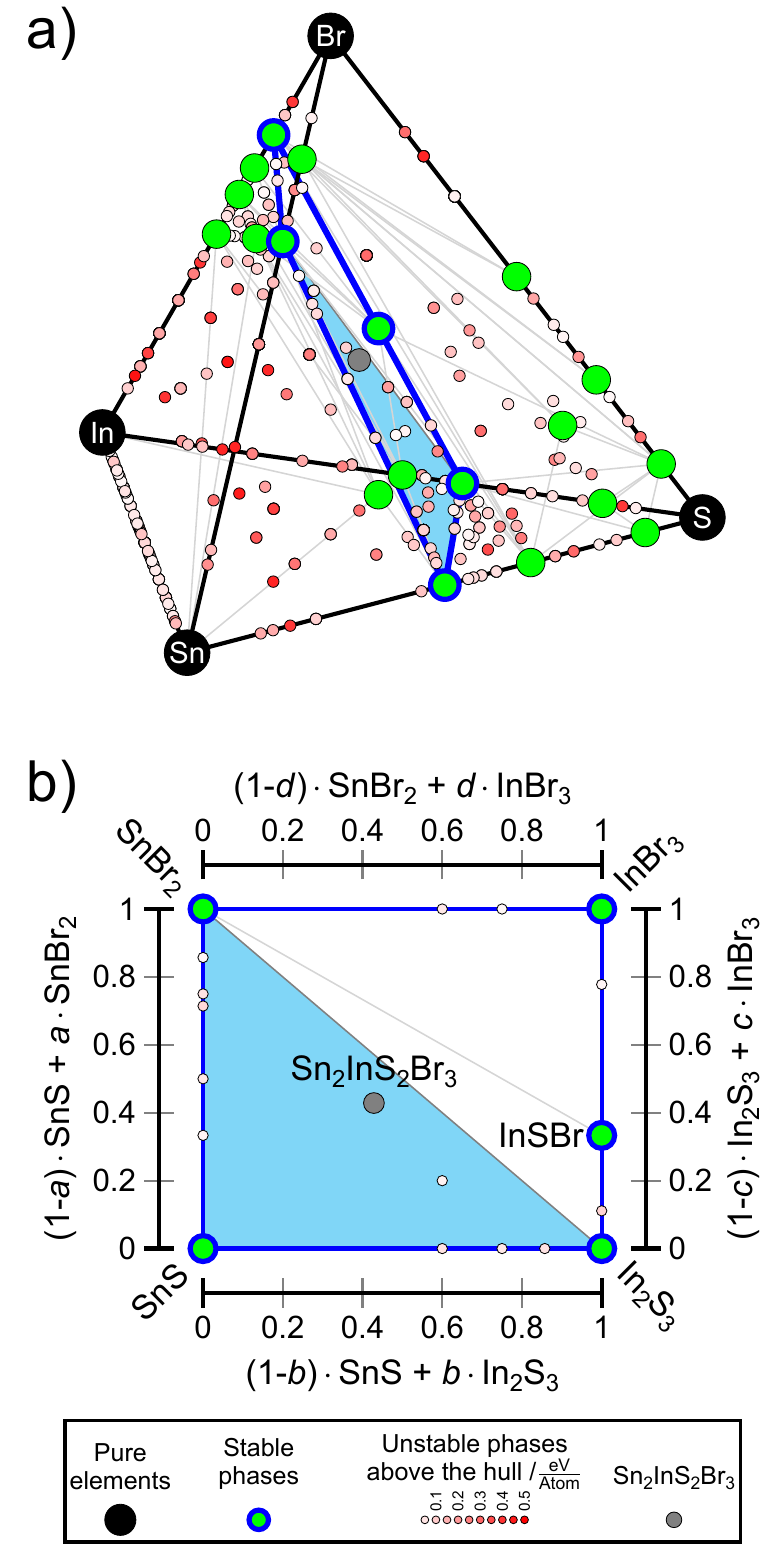}
\caption{Compositional phase diagram of \ch{Sn2InS2Br3} including all competing phases with energies $E_\text{hull} \leq \SI{0.5}{\electronvolt\per\at}$. a) Full four-dimensional phase diagram, and b) two-dimensional surface projection (blue lines) illustrating the convex hull region (cyan-coloured triangle) containing the respective MMCH compounds (here \ch{Sn2InS2Br3}). Stable compounds highlighted with a blue frame (\ch{SnBr2}, \ch{InBr3}, \ch{InSBr}, \ch{In2S3}, and \ch{SnS}) define the vertices of the 2D projection shown in b).}
\label{fig:phasediagram_Sn2InS2Br3}
\end{center}
\end{figure}

The calculated compositional phase diagrams, as illustrated for \ch{Sn2InS2Br3} in Figure\,\ref{fig:phasediagram_Sn2InS2Br3}, consistently illustrate that all 54 MMCH compounds would decompose into competing phases, irrespective of the space group. Figure\,\ref{fig:phasediagram_Sn2InS2Br3}\,a) exemplarily displays the four-dimensional compositional phase diagram of \ch{Sn2InS2Br3}, which, owing to the quaternary nature of MMCH compounds, comprises a large number of competing phases. Within this compositional phase diagram, \ch{Sn2InS2Br3} (gray point) is located close to the centre. Analysing the compositional phase diagram reveals that \ch{Sn2InS2Br3} decomposes into \ch{In2S3}, \ch{SnS}, and \ch{SnBr2}. These phases lie within a two-dimensional subspace spanned by \ch{In2S3}, \ch{SnS}, \ch{SnBr2}, and \ch{InBr3}. This subspace is obtained by projecting the four-dimensional phase space onto the plane defined by these corner phases, as shown in Figure\,\ref{fig:phasediagram_Sn2InS2Br3}\,b), with details on the projection procedure provided in SI Section\,S3. Within this plane, \ch{Sn2InS2Br3} is positioned near the centre but slightly shifted toward the south-west in the direction of \ch{SnS}, while additional unstable phases are shown as red-shaded points according to their hull energies ($E_\text{hull}$). The three stable decomposition products define a convex hull surface (cyan-highlighted), within which \ch{Sn2InS2Br3} is located, see Figure\,\ref{fig:phasediagram_Sn2InS2Br3}\,b). This representation enables direct identification of the decomposition products of the respective MMCH compounds.

Overall, despite being predicted to be unstable, the compositional phase diagrams provide a consistent framework for analysing and comparing the decomposition pathways of MMCH compounds.

\subsection{2.2 Decomposition Reactions}
Despite spanning a chemically diverse space, computed compositional phase diagrams for all 54 studied MMCHs reveal that they decompose through only five distinct reaction pathways:

\begin{align}
	\begin{split}
	\label{eq:reaction_1}
	\textbf{R1:} \quad &2\,\ch{M(II)2M(III)Ch2X3} \leftrightarrow \ch{M(III)_2Ch_3}\\
    &+ 3\,\ch{M(II)X_2} + \ch{M(II)Ch}
	\end{split}\\
	\begin{split}
	\label{eq:reaction_2}
	\textbf{R2:} \quad &\ch{M(II)2M(III)Ch2X3} \leftrightarrow \ch{M(III)ChX}\\
    &+ \ch{M(II)X_2} + \ch{M(II)Ch}
	\end{split}\\
	\begin{split}
	\label{eq:reaction_3}
	\textbf{R3:} \quad &2\,\ch{M(II)2M(III)Ch2X3} \leftrightarrow \ch{M(II)M(III)_2Ch_4}\\
    &+ 3\,\ch{M(II)X_2} 
	\end{split}\\
	\begin{split}
	\label{eq:reaction_4}
	\textbf{R4:} \quad &6\,\ch{M(II)2M(III)Ch2X3} \leftrightarrow \ch{M(II)M(III)_6Ch_{10}}\\
    &+ 9\,\ch{M(II)X_2} + 2\,\ch{M(II)Ch} 
	\end{split}\\
	\begin{split}
	\label{eq:reaction_5}
	\textbf{R5:} \quad &8\,\ch{M(II)2M(III)Ch2X3} \leftrightarrow \ch{M(II)_3M(III)_8Ch_{15}}\\
    &+ 12\,\ch{M(II)X_2} + \ch{M(II)Ch}
	\end{split}
\end{align}%
A comprehensive overview of the decomposition behaviour of all 54 MMCH compounds is provided in SI Table\,S1. Overall, reactions\,(\ref{eq:reaction_4}) and (\ref{eq:reaction_5}) occur exclusively for \ch{Pb}-based MMCH compounds, whereas reaction\,(\ref{eq:reaction_2}) is observed solely for \ch{Sn}-based MMCHs. \ch{Sn2InS2Br3} represents an example of decomposition following reaction\,(\ref{eq:reaction_1}) (see Figure \ref{fig:phasediagram_Sn2InS2Br3} b)), while \ch{Sn2InTe2I3}, \ch{Sn2SbTe2I3}, \ch{Pb2InTe2I3}, and \ch{Pb2SbS2I3} decompose through reactions\,(\ref{eq:reaction_2}), (\ref{eq:reaction_3}), (\ref{eq:reaction_4}) and (\ref{eq:reaction_5}), respectively, see SI Figures\,S2 and S3. The decomposition pathway of a given MMCH compound is primarily determined by the \ch{M(II)}, \ch{M(III)}, and \ch{Ch} sites, while the \ch{X} site has almost no influence. All three halogens exhibit the same decomposition reactions for \ch{Pb}-based MMCHs and for nearly all \ch{Sn}-based counterparts. Exceptions occur for \ch{Sn2BiSe2X3}, where \ch{Br} and \ch{I} decompose \textit{via} reaction\,(\ref{eq:reaction_2}) but \ch{Cl} follows (\ref{eq:reaction_1}), and for \ch{Sn2BiTe2X3}, where \ch{Cl} and \ch{I} decompose \textit{via} reaction\,(\ref{eq:reaction_3}) while \ch{Br} follows (\ref{eq:reaction_2}).

\begin{figure}
\begin{center}
\includegraphics[width=0.5\textwidth]{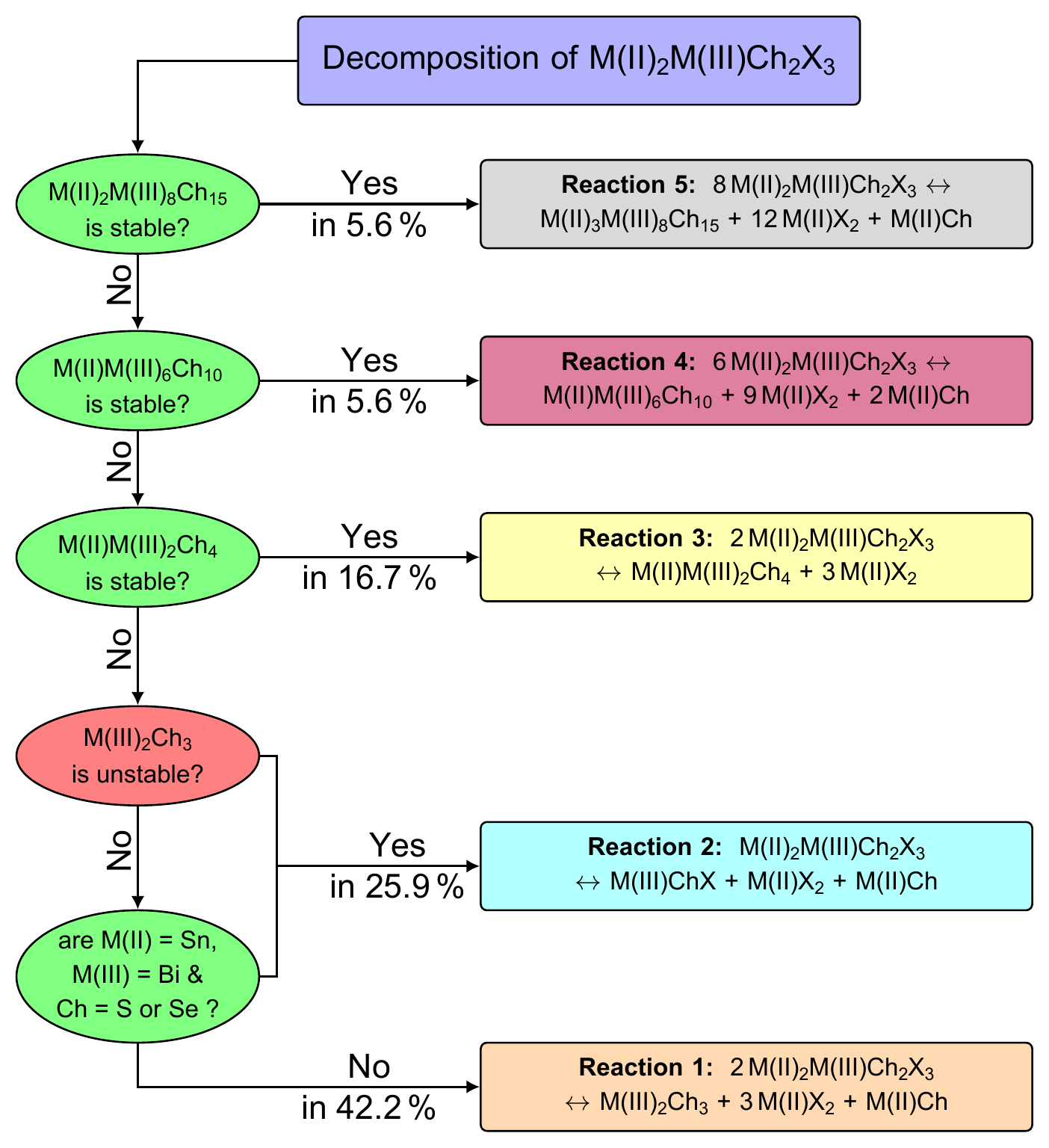}
\caption{Decomposition reactions of \ch{M(II)2M(III)Ch2X3} compounds, along with the conditions under which each reaction occurs. Percentages indicate the proportion of the 54 MMCH compounds that follow each respective decomposition reaction.}
\label{fig:decomp_recept}
\end{center}
\end{figure}

The calculated phase diagrams and corresponding decomposition pathways\, (\ref{eq:reaction_1}) to (\ref{eq:reaction_5}) enable the formulation of a general \enquote{recipe} for MMCH decomposition. This framework is primarily governed by the relative stability of competing ternary and key binary phases, as illustrated in reactions\,(\ref{eq:reaction_5}) to (\ref{eq:reaction_3}) and in reactions\,(\ref{eq:reaction_2}) to (\ref{eq:reaction_1}), respectively. More precisely, a set of targeted questions determines which decomposition pathway a given MMCH compound follows, see Figure\,\ref{fig:decomp_recept}.

Reaction\,(\ref{eq:reaction_1}) is the default decomposition pathway and is observed for \SI{42.2}{\percent} of all investigated MMCH compounds. Decomposition deviates from this pathway for three reasons: A) stable IV-V-VI mixed-metal chalcogenides form on the line between the binary metal chalcogenides \ch{M(II)Ch} and \ch{M(III)2Ch3}, B) the corresponding \ch{M(III)2Ch3} phase is unstable, or C) a competing decomposition reaction suppresses reaction\,(\ref{eq:reaction_1}).

The first exception arises when a stable IV-V-VI mixed-metal chalcogenide replaces one or both binary metal chalcogenides decomposition products. If \ch{M(II)M(III)_2Ch_4} (adopting the rhombohedral \ch{GeAs2Te4} structure type),\cite{wang2022, sergeev2025, mcguire2023} is stable, decomposition proceeds via reaction\,(\ref{eq:reaction_3}). Stable representatives include the \ch{PbBi2Ch4} family (Ch = \ch{S}, \ch{Se}, and \ch{Te}),\cite{cech1978, peacock1940, agaev1968, comodi2019, chatterjee2014, shvets2017} making this the predominant decomposition pathway for \ch{Pb2BiCh2X3} compounds. Additionally, stable \ch{Sn}-based tellurides, namely \ch{SnSb2Te4}\cite{talybov1961} and \ch{SnBi2Te4},\cite{adouby2000} likewise favour this pathway for \ch{Sn2SbTe2X3} and \ch{Sn2BiTe2X3} compounds, except for \ch{Sn2BiTe2Br3}, which decomposes \textit{via} reaction\,(\ref{eq:reaction_2}). If \ch{M(II)M(III)_6Ch_{10}} is stable, decomposition follows reaction\,(\ref{eq:reaction_4}). So far, \ch{M(II)M(III)_6Ch_{10}} has only been predicted computationally for \ch{PbIn_{6}Te_{10}}, making this pathway unique to \ch{Pb2InTe2X3}. Finally, if the fülöppite-type compound \ch{M(II)_2M(III)_8Ch_{15}} is stable, decomposition proceeds via reaction,(\ref{eq:reaction_5}). Since this phase is only known for \ch{Pb3Sb8S15},\cite{edenharter1974} this pathway is exclusive to \ch{Pb2SbS2X3}.

The second exception occurs when the corresponding \ch{M(III)2Ch3} phase is unstable, resulting in decomposition \textit{via} reaction~(\ref{eq:reaction_2}). This is the case for all \ch{M(II)2InTe2X3} compounds. However, because the \ch{Pb}-based compounds decompose \textit{via} reaction~(\ref{eq:reaction_4}), only \ch{Sn2InTe2X3} follow this pathway.

The third exception is caused by a competing decomposition reaction yielding \ch{Bi2Ch3}, \ch{SnCh}, and \ch{BiChX} becomes thermodynamically favourable, causing MMCHs to decompose \textit{via} reaction\,(\ref{eq:reaction_2}), see SI Figure\,S3\,b). This occurs for all \ch{Sn2BiS2X3} and \ch{Sn2BiSe2X3} compounds except \ch{Sn2BiSe2Cl3}, which instead decomposes \textit{via} reaction\,(\ref{eq:reaction_1}).

The decomposition behaviour of \ch{Pb}- and \ch{Sn}-based MMCH compounds differs significantly as a consequence of the stability of the corresponding decomposition products. No stable \ch{Sn2M(III)_{8}Ch_{15}} or \ch{SnM(III)_{6}Ch_{10}} compounds have been reported to date, indicating that \ch{Sn}-based MMCH compounds decompose exclusively according to reactions\,(\ref{eq:reaction_1}) to (\ref{eq:reaction_3}). Decomposition according to reaction\,(\ref{eq:reaction_2}) is not observed for \ch{Pb}-based MMCHs, which can be explained by the high stability of \ch{Pb}- and chalcogenide-based decomposition phases such as \ch{PbIn_{6}Te_{10}}, \ch{PbBi2S4}, and \ch{PbBi2Se4}. The stability of \ch{PbIn_{6}Te_{10}} causes \ch{Pb2InTe2X3} compounds to decompose \textit{via} (\ref{eq:reaction_4}), despite the instability of \ch{In2Te3}, even though reaction (\ref{eq:reaction_2}) would be expected by analogy with the \ch{Sn}-based counterparts. An analogous behaviour is observed for \ch{PbBi2S4} and \ch{PbBi2Se4}, which are stable in contrast to their \ch{Sn}-based counterparts. Consequently, \ch{Pb2BiS2X3} and \ch{Pb2BiSe2X3} decompose \textit{via} reaction\,(\ref{eq:reaction_3}) rather than (\ref{eq:reaction_2}).

Overall, the decomposition behaviour of all 54 MMCH compounds can be condensed into only five reaction pathways that are governed primarily by the stability of competing binary phases and, to a lesser extent, ternary phases.

\subsection{2.3 Hull Energies}
The energy above the convex hull, $E_\text{hull}$, quantifies the thermodynamic driving force for decomposition and is defined as the distance of a compound from the convex hull, corresponding to the energy difference to the tie-line (or facet) defined by the lowest-energy combination of stable phases. Smaller $E_\text{hull}$ values indicate a closer proximity to stability and thus a higher likelihood that a compound may be experimentally realisable, whereas larger values imply a stronger tendency to decompose into competing phases. It should be noted that $E_\text{hull}$ does not capture kinetic barriers; therefore, no conclusions about decomposition kinetics or likelihood of MMCHs can be drawn.

\begin{figure}
	\centering
	\includegraphics[width=0.75\textwidth]{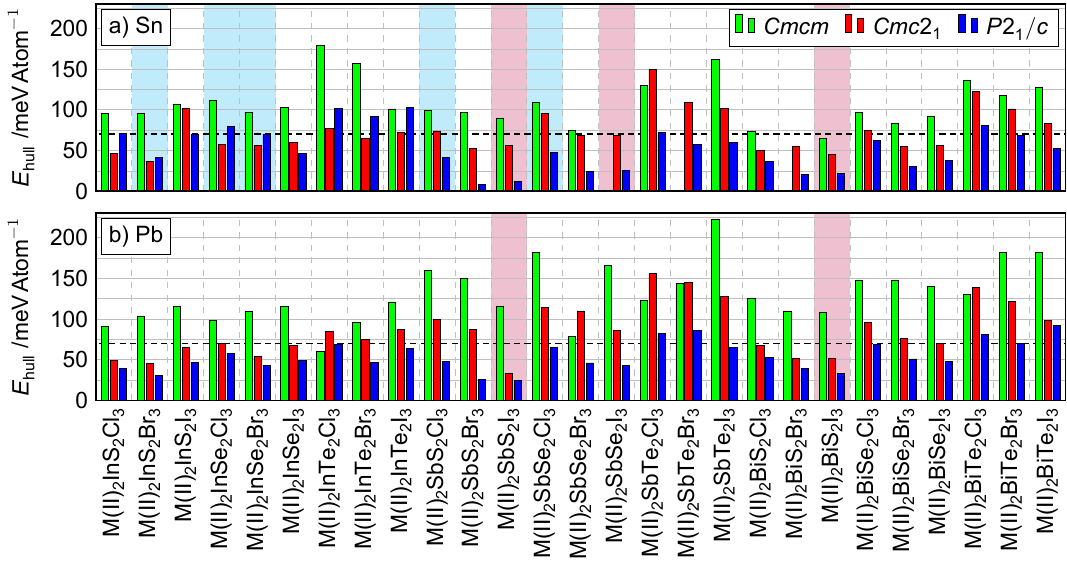}
	\caption{Hull energies (PBE level of theory) for a) tin and b) lead MMCHs Colours: Black: estimated thermodynamic stability uncertainty derived from Bartel \textit{et al.}\cite{bartel2019}; Purple: MMCH compounds that have been previously studied experimentally and/or theoretically in the literature;\cite{nie2020b, kavanagh2021b, olivierfourcade1980, nicolson2023, roth2024, ibanez1984, islam2016, dolgikh1985, doussier2007, roth2023}  Cyan: lead-free MMCH compounds experimentally investigated in this work, see SI Section\,S2 for experimental details and SI Section\,S7 for the corresponding powder X-ray diffraction patterns and Rietveld refinements.}
	\label{fig:hull_energies}
\end{figure}

Figure\,\ref{fig:hull_energies} shows the hull energies for all 54 materials. It reveals three key trends. Firstly, all 54 investigated MMCH compounds exhibit $E_\text{hull} > 0$, see also SI Table\,S1 for a detailed listing. The positive hull energies indicate thermodynamic instability with respect to phase separation, with decomposition (see reactions\,(\ref{eq:reaction_1})-(\ref{eq:reaction_5})) into the corresponding products being energetically favoured. Secondly, across all phases, \ch{Sn}-based MMCHs generally lie closer to the convex hull than their \ch{Pb}-based counterparts. The average hull energies of \ch{Sn}-based compounds are \SI[separate-uncertainty=true]{108(27)}{}, \SI[separate-uncertainty=true]{74(26)}{}, and \SI[separate-uncertainty=true]{53(21)}{\milli\electronvolt\per\at} for the $Cmcm$, $Cmc2_1$, and $P2_1/c$ phases, compared with \SI[separate-uncertainty=true]{131(36)}{}, \SI[separate-uncertainty=true]{86(32)}{}, and \SI[separate-uncertainty=true]{55(18)}{\milli\electronvolt\per\at} for the corresponding \ch{Pb}-based compounds. Thirdly, in nearly all cases, the $P2_1/c$ phase is the most stable polymorph, except for \ch{Sn2InS2Cl3}, \ch{Pb2InTe2Cl3}, \ch{Sn2InTe2I3}, and \ch{Sn2InTe2Cl3}. On average, compounds in the $P2_1/c$ phase lie \SI[separate-uncertainty=true]{54(22)}{\milli\electronvolt\per\at} above the convex hull, followed by the $Cmc2_1$ phase with \SI[separate-uncertainty=true]{80(30)}{\milli\electronvolt\per\at} and the $Cmcm$ phase with \SI[separate-uncertainty=true]{120(34)}{\milli\electronvolt\per\at}. 

In addition, our RF and SHAP analysis further corroborated the enhanced stability of \ch{Sn}-based MMCHs relative to their \ch{Pb}-based counterparts, see SI Figures\,S4-S6. Independent of the space group, the presence of \ch{Sn} consistently reduces $E_\text{hull}$. Moreover, the SHAP analysis reveals that the \ch{Ch} site has the largest impact on the hull energy, with \ch{S} and \ch{Se} reducing $E_\text{hull}$ (\ch{S} more than \ch{Se}) and \ch{Te} increasing it. Following the chalcogen site, the \ch{M(III)} site has the second largest influence on the hull energy, while the local contributions of \ch{In}, \ch{Sb}, and \ch{Bi} depend on the specific space group, see SI Table\,S1. In contrast, the influence of the halogens is fully determined by the space group, see SI Figures\,S4-S6 and Table\,S3.

The prediction that \ch{Sn}-based MMCHs are more stable than their \ch{Pb}-based counterparts is surprising for several reasons. Firstly, our previous studies based on stability versus elemental decomposition showed the opposite trend, with \ch{Pb}-based MMCHs being significantly more stable than their \ch{Sn}-based analogues.\cite{henkel2023, henkel2025}  Secondly, the enhanced stability of \ch{Sn}-based MMCHs contradicts trends observed in perovskites, where \ch{Pb}-based compounds are generally reported to be more stable than their \ch{Sn}-based counterparts.\cite{kaiser2022, pascual2020, akbulatov2019, leijtens2017} Thirdly, from a chemical perspective, this apparent contradiction is further underscored by the well-known tendency of tin to oxidise more readily from \ch{Sn^{2+}} to \ch{Sn^{4+}} than lead.\cite{he2024} The origin of this discrepancy can be rationalised by considering how the hull energy depends on the stability of competing decomposition phases. Whether a compound lies on or above the convex hull is not only determined by its own formation energy, but crucially by the energies of the most stable neighbouring phases/decomposition products. Analysing the formation energies of the decomposition products (see reactions\,(\ref{eq:reaction_1}) to (\ref{eq:reaction_5})) reveals a clear trend: \ch{Sn}-based decomposition products (mean: \SI{-577}{\milli\electronvolt\per\fu}) are energetically less favourable  than their \ch{Pb}-based counterparts (mean: \SI{-684}{\milli\electronvolt\per\fu}), while \ch{M(III)} decomposition products have intermediate energies (mean: \SI{-598}{\milli\electronvolt\per\fu}), see SI Figure\,S7. In contrast, the formation energies of \ch{Sn}- and \ch{Pb}-based MMCHs are similar, with a mean difference of only $\sim\SI{70}{\milli\electronvolt\per\fu}$ across all three space groups. The distance to the convex hull is therefore primarily governed by the stability of the competing decomposition products. Since \ch{Sn}-based decomposition phases are energetically less favourable than their \ch{Pb}-counterparts, \ch{Sn}-based MMCHs are shifted closer to the convex hull and thus exhibit systematically lower energies. 

All 54 MMCH compounds exhibit $E_\text{hull} > 0$, with $P2_1/c$ structures lying closer to the convex hull and thus being more stable than the $Cmc2_1$ and $Cmcm$ ones. The observed stability trend $P2_1/c$ > $Cmc2_1$ > $Cmcm$ is in good agreement with our previous studies on MMCH compounds based on formation energies.\cite{henkel2023, henkel2025} It is not surprising that the $P2_1/c$ phase emerges as the most stable structure, as it represents a low-temperature phase. Doussier \textit{et al.} demonstrated for \ch{Pb2SbS2I3} that the orthorhombic $Cmcm$ phase, which is stable at room temperature, undergoes a transition to the monoclinic $P2_1/c$ phase at below \SI{100}{\kelvin}.\cite{doussier2007} Consequently, this phase naturally dominates our DFT-based hull energy calculations performed at \SI{0}{\kelvin}. In addition, Kavanagh \textit{et al.} showed on the basis of DFT calculations that the $Cmcm$ phase of \ch{Sn2SbS2I3} is unstable at \SI{0}{\kelvin}, as evidenced by imaginary phonon modes and can be interpreted as an average over the energetically more favourable, lower-symmetry $Cmc2_1$ configurations.\cite{kavanagh2021b} Accordingly, we predict the $Cmcm$ phase to be the least stable, despite experimental evidence indicating that it is the preferred phase at room temperature.\cite{islam2016,olivierfourcade1980, ibanez1984, doussier2007, roth2024} 
The $Cmc2_1$ phase can therefore be regarded as an intermediate between the experimentally observed room-temperature $Cmcm$ phase and the low-temperature $P2_1/c$ phase. Given the \SI{0}{\kelvin} nature of DFT calculations,  particular relevance can be assigned to the $Cmc2_1$ phase when assessing the stability landscape of MMCHs.

Interpreting convex hull analyses derived from DFT requires careful consideration of uncertainties in the underlying exchange-correlation (XC) functional and of thermodynamic effects that influence the link between computational predictions and experimental phase stability, as previous studies have shown that the quantitative agreement between predicted and measured hull energies strongly depends on both the material class and the chosen XC functional.\cite{bartel2019, hautier2012} In this context, Hautier \textit{et al.} analysed \num{135} experimentally known ternary oxides including systems with transition metals and reported a standard deviation of \SI{24}{\milli\electronvolt\per\at} for hull energies obtained using PBE+U.\cite{hautier2012} By contrast, Bartel \textit{et al.} examined a substantially broader dataset of \num{1012} inorganic solids including oxides, chalcogenides, and halides as well as systems containing transition metals, and reported mean absolute deviations (MADs) of \SI{70}{\milli\electronvolt\per\at} for PBE and \SI{59}{\milli\electronvolt\per\at} for SCAN.\cite{bartel2019} While quaternary MMCHs were not explicitly covered in either study, the materials considered by Bartel \textit{et al.} include both halogen- and chalcogen-containing compounds, making the reported MAD of \SI{70}{\milli\electronvolt\per\at} for PBE a reasonable and conservative lower bound for the uncertainty in the present work. 

Accordingly, we adopted \SI{70}{\milli\electronvolt\per\at} as the uncertainty threshold in Figure\,\ref{fig:hull_energies} (black dotted line). Considering all structures within \SI{70}{\milli\electronvolt\per\at} of the convex hull as potentially stable, the stability landscape is substantially less restrictive than suggested by a strict hull energy criterion. For the $Cmcm$ space group, consistent with its instability at \SI{0}{\kelvin}, only \ch{Sn2BiSe2Cl3} and \ch{Pb2InTe2Cl3} fall within this threshold, while all other compounds are above it. In contrast, \num{26} MMCHs (\num{15} \ch{Sn}-based and \num{11} \ch{Pb}-based) in $Cmc2_1$ and \num{43} MMCHs (\num{20} \ch{Sn}-based and \num{23} \ch{Pb}-based) in $P2_1/c$ are within the threshold and can therefore be regarded as potentially thermodynamically stable. Notably, all five experimentally reported MMCHs \-- \ch{Sn2SbS2I3}, \ch{Sn2SbSe2I3}, \ch{Sn2BiS2I3}, \ch{Pb2SbS2I3}, and \ch{Pb2BiS2I3} (highlighted in purple in Figure\,\ref{fig:hull_energies} and listed in SI Table\,S1) \-- fall below this threshold in at least the $Cmc2_1$ and $P2_1/c$ space groups. Considering that these five materials have been made provides confidence in our adopted uncertainty criterion and suggests that other MMCHs whose hull energies fall within the \SI{70}{\milli\electronvolt\per\at} may also be synthesisable.

Overall, the thermodynamic accessibility of MMCH compounds, reflected in their proximity to the convex hull, is governed primarily by two factors. First, the composition, particularly the \ch{M(II)} and \ch{Ch} site elements, with \ch{Sn}-based compounds generally being more stable than their \ch{Pb}-counterparts. Second, the crystal structure, with the low-symmetry $P2_1/c$ phase being consistently favoured.


\subsection{2.4 Synthesis of MMCH compounds}
We investigated six \ch{Sn}-based MMCH compounds using solution processing and X-ray diffraction (XRD)/Rietveld refinement to assess whether the computed compositional phase diagrams capture experimentally relevant limits on synthetic accessibility. Further methodological details are provided in SI Section\,S2. The selection was guided by the aim to cover a broad range of \ch{Sn}-based compounds with low predicted $E_\text{hull}$ values for the room temperature relevant $Cmcm$ and $Cmc2_1$ phases, while avoiding the inclusion of heavy metals like \ch{Pb} or \ch{Bi}.  The focus on Sn-based compounds was further motivated by the predicted high defect tolerance of \ch{Sn2SbS2I3}.\cite{kavanagh2021b,nicolson2023} Moreover, our previous study on the influence of different substitutional elements on the MMCHs material properties showed that, among the chalcogens, \ch{S} and \ch{Se}, as well as \ch{Cl} and \ch{Br} among the halogens, have a particularly beneficial effect on the formation energy, band gap, and charge-carrier effective masses.\cite{henkel2025} Therefore, we chose \ch{Sn2InS2Br3}, \ch{Sn2InSe2Br3}, \ch{Sn2SbS2Cl3}, \ch{Sn2SbSe2Cl3}, and \ch{Sn2InSe2Cl3}, along with \ch{Sn2SbS2I3} as a reference material, whose successful synthesis with high phase purity has recently been reported by Nie \textit{et al.}\cite{nie2020a}. According to our computations (see Figure\,\ref{fig:decomp_recept}), all six MMCHs are predicted to decompose via reaction\,(\ref{eq:reaction_1}). Further details on the broader experimental screening campaign and the XRD-derived phase identification across all samples are provided in SI Section\,S7 and SI Table\,S1, respectively.

\begin{figure}
\begin{center}
\includegraphics[width=0.75\textwidth]{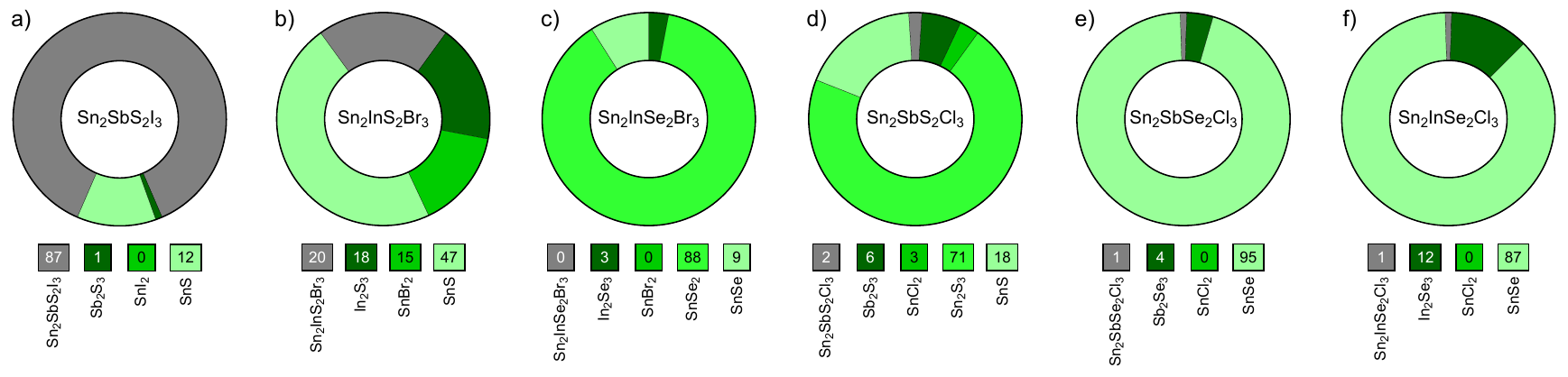}
\caption{Pie charts of the normalised phase composition (excluding contributions from the substrate and possible oxides) for a) \protect\ch{Sn2SbS2I3}, b) \protect\ch{Sn2InS2Br3}, c) \protect\ch{Sn2InSe2Br3}, d) \protect\ch{Sn2SbS2Cl3}, e) \protect\ch{Sn2SbSe2Cl3}, and f) \protect\ch{Sn2InSe2Cl3}, as determined by Rietveld refinement, see SI Figure\,S8 for the corresponding powder X-ray diffraction patterns and SI Table\,S3 for all phase compositions, including the substrate and possible oxides. Colours: gray corresponds to the respective MMCH compound, while shades of green denotes decomposition products.}
\label{fig:phase_comp}
\end{center}
\end{figure}

Figure\,\ref{fig:phase_comp} summarises the phase composition of the six MMCH compounds based on the performed refinements. A comprehensive list of all phases included in the refinements, encompassing potential oxides, synthesis by-products, and substrate-related phases, is given in SI Table\,S3. The results indicate that \ch{Sn2SbS2I3} can be synthesised with high phase purity, whereas the hitherto unreported \ch{Sn2InS2Br3} (CCDC database entry No. 2553969\cite{ccdc_sn2sbs2i3}) is formed in noticeable, albeit limited, quantities. For \ch{Sn2SbS2Cl3}, the observed phase composition suggests transient formation followed by rapid decomposition. In contrast, the syntheses of \ch{Sn2InSe2Br3}, \ch{Sn2SbSe2Cl3}, and \ch{Sn2InSe2Cl3} were not successful.

Our computational and experimental results indicate that MMCH phase formation is governed by strong thermodynamic driving forces toward binary decomposition products. Thus, the experimental data can be interpreted in two ways: either MMCH phases form only transiently and subsequently decompose, or competing binary and ternary phases nucleate preferentially, effectively suppressing the formation of phase-pure MMCH compounds.

A common feature observed across all six compounds is the pronounced formation of tin chalcogenide phases, including \ch{Sn^{2+}} compounds such as \ch{SnS} and \ch{SnSe}, as well as \ch{Sn^{4+}}-containing derivatives such as \ch{SnSe2} and \ch{Sn2S3}. The formation of these phases is at least in part due to decomposition, most notably for \ch{Sn2InS2Br3} and, to a lesser extent, \ch{Sn2SbS2Cl3}. However, they are also well-known synthesis by-products arising from the use of thiourea (TU) and selenourea precursors, as used in this study. For instance, Rao \textit{et al.} demonstrated that \ch{SnCl2}(aq) reacts with TU under hydrothermal conditions to form \ch{SnS} nanostructures,\cite{rao2004} while Brune \textit{et al.} showed that $N,N$-dimethylselenourea can act as a single-source precursor for \ch{SnSe} nanoparticles \textit{via} defined Sn-chalcogen complexes.\cite{brune2021} Therefore, the occurrence of chalcogenides should not be over-interpreted as proof of MMCH decomposition, as they may instead reflect competition between MMCH formation and early by-products resembling its predicted decomposition products. 

The successful and highly phase-pure synthesis of \ch{Sn2SbS2I3} is in excellent agreement with previous experimental studies by Olivier-Fourcade \textit{et al.},\cite{olivierfourcade1980} Nie \textit{et al.},\cite{nie2020b} Roth \textit{et al.},\cite{roth2024} and Pratolongo \textit{et al.}\cite{pratolongo2025} Notably, essentially no decomposition into \ch{Sb2S3}, \ch{SnI2}, or \ch{SnS} is observed. Only \ch{SnS} is detected, contributing approximately \SI{12}{\percent} to the overall phase composition, see Figure\,\ref{fig:phase_comp} a). Additionally, a minor amount of \ch{SnO2} (\SI{2.5}{\percent} in absolute terms, see SI Table\,S3) is present, which may indicate slight oxidation of \ch{Sn^{2+}} to \ch{Sn^{4+}} or substrate-related effects. Our results confirm that \ch{Sn2SbS2I3} can be obtained in high phase purity under solution-processing conditions.

\ch{Sn2InS2Br3} falls within an intermediate regime, as the Rietveld refinement clearly shows the coexistence of the target MMCH phase ($\sim \SI{20}{\percent}$) with its predicted decomposition products \ch{In2S3}, \ch{SnBr2}, and \ch{SnS} (see Figure\,\ref{fig:phase_comp} b)), and is therefore in full agreement with reaction\,(\ref{eq:reaction_1}). As for the other compounds, \ch{SnS} dominates the phase composition, explaining deviations from the ideal stoichiometric ratios expected from reaction\,(\ref{eq:reaction_1}). Nevertheless, the successful (partial) formation of \ch{Sn2InS2Br3} demonstrates that this compound can be accessed transiently or partially under optimised conditions and simultaneously validates the predicted decomposition pathway. A repeated synthesis of \ch{Sn2InS2Br3} produced similar results, with the refinement indicating a phase contribution of $\sim \SI{30}{\percent}$. However, this sample (No. 4, see SI Table\,S1) exhibited pronounced oxidation, resulting in the formation of about \SI{15}{\percent} combined \ch{SnO2} and \ch{SnS2} impurities. For \ch{Sn2SbS2Cl3} the selected pattern is dominated by \ch{Sn2S3} and \ch{SnS}, with only minor contributions from \ch{Sb2S3} and \ch{SnCl2}, see Figure\,\ref{fig:phase_comp} d). Although inclusion of the target structural model accommodates a small fraction of the fitted intensity, its estimated contribution of $\sim \SI{2}{\percent}$ is comparable to the practical uncertainty of the multiphase thin-film refinement. The refinement data therefore do not convincingly resolve \ch{Sn2SbS2Cl3} and do not establish either transient target formation or rapid target decomposition. However, the refined phase composition is dominated by tin chalcogenides, especially \ch{Sn2S3}, a mixed-valence compound containing both \ch{Sn^{2+}} and \ch{Sn^{4+}}. Therefore, the phase formation of \ch{Sn2SbS2Cl3} is not governed by a single mechanism. Instead, the computed compositional phase diagram favours the formation of secondary binary phases, while \ch{Sn} oxidation oxidation further hinder the stabilisation of the quaternary phase in phase-pure form. Overall, the cases of \ch{Sn2InS2Br3} and \ch{Sn2SbS2Cl3} highlight a strong agreement between theoretical predictions and experimental observations for the relevant decomposition paths.

In contrast, the three \ch{Se} compositions \ch{Sn2InSe2Br3}, \ch{Sn2SbSe2Cl3}, and \ch{Sn2InSe2Cl3} could not be successfully synthesised under the employed conditions. Accordingly, the refinements for \ch{Sn2InSe2Br3} (see Figure\,\ref{fig:phase_comp} c)), \ch{Sn2SbSe2Cl3} (see Figure\,\ref{fig:phase_comp} e)), and \ch{Sn2InSe2Cl3} (see Figure\,\ref{fig:phase_comp} f)) show no evidence of the corresponding MMCH phases. Instead, all three samples are dominated by tin chalcogenide derivatives, with MMCH contributions remaining $\lesssim \SI{1}{\percent}$. In the case of \ch{Sn2InSe2Br3}, a dominant contribution of \ch{SnSe2} ($\sim \SI{88}{\percent}$) is observed, indicating extensive oxidation of \ch{Sn^{2+}} to \ch{Sn^{4+}}. By contrast, \ch{Sn2SbSe2Cl3} and \ch{Sn2InSe2Cl3} are primarily composed of \ch{Sn^{2+}} chalcogenides, with no evidence of significant oxidation processes. Overall, the absence of \ch{SnX2} species and the corresponding unsuccessful incorporation of halogens, together with the negligible MMCH phase contributions, strongly suggest that the synthesis pathway appears to be dominated by direct nucleation of energetically more favourable binary phases.

Regarding halogen incorporation into crystalline (MMCH) phases, our experimental results reveal a strong halogen dependence. Iodine was retained within the dominant \ch{Sn2SbS2I3} phase, whereas \ch{Br}- and \ch{Cl}-containing crystalline phases are detected less consistently.  The limited retention of \ch{Cl} and \ch{Br} may reflect halide loss, incorporation into amorphous or poorly crystalline material, redistribution among precursor-derived species, or concentrations below the XRD detection limit, likely resulting from halide loss and redistribution during processing due to their high volatility,\cite{holleman2007} These effects may account for the limited observation of \ch{SnX2} phases across the six investigated MMCH compounds, despite their predicted occurrence in reaction\,(\ref{eq:reaction_1}). Overall, the experimental results suggest that halide incorporation and retention are important factors in MMCH phase formation under the investigated processing conditions, with iodine- and, to a lesser extent, bromide-based compounds being more readily accessible than chloride-rich analogues.

These insights are further supported by our extensive experimental screening campaign, see SI Table\,S1 and SI Section\,S7. This is particularly evident for \ch{Sn2M(III)Ch2Cl3} MMCHs and \ch{Sn2InSe2Br3}. Across a wide range of synthesis parameters, including variations in precursor ratios, annealing temperature, precursor chemistry (e.g., acetate precursors), addition of metallic \ch{Sn}, substrate choice, and alternative reaction pathways such as \ch{SnCh} + \ch{SbCl3}, the reactions consistently yielded binary (decomposition) products and halide-deficient phases.

Our experimental insights reveal a clear trend in the synthetic accessibility of MMCHs; while iodide MMCHs such as \ch{Sn2SbS2I3} can be obtained in high phase purity, bromide systems show partial phase formation, and \ch{Cl} and \ch{Se} compounds systematically fail to form under all investigated reaction conditions. Across nearly all MMCH compounds, phase formation is dominated by the rapid nucleation of thermodynamically stable binary phases, most prominently tin chalcogenides, in combination with \ch{Sn^{2+}} oxidation toward \ch{Sn^{4+}} species and poor incorporation of halides. These effects collectively outcompete the formation of quaternary MMCH phases, even when alternative synthesis routes and parameter variations are employed.

Overall, these observations strongly support the predicted thermodynamic limitations and demonstrate that the practical synthesis of MMCHs is governed by kinetic competition and chemical stability constraints.

\section{3 Conclusion}
This study sets out to evaluate the thermodynamic accessibility of quaternary MMCHs and reveals that the main obstacle to their solution synthesis is not the absence of viable precursor routes but the strong thermodynamic competition from low‑energy binary and ternary phases. By computing and analysing the compositional phase diagrams of 54 MMCH compounds in three space groups ($Cmcm$, $Cmc2_1$, and $P2_1/c$), determining their corresponding hull energies, and examining the synthetic accessibility of six selected Sn‑based MMCH compounds (\ch{Sn2SbS2I3}, \ch{Sn2InS2Br3}, \ch{Sn2InSe2Br3}, \ch{Sn2SbS2Cl3}, \ch{Sn2SbSe2Cl3}, and \ch{Sn2InSe2Cl3}) using a combined computational–experimental approach, we find that all 54 MMCH compounds lie above the convex hull at \SI{0}{\kelvin}, indicating limited thermodynamic accessibility. Accounting for computational uncertainty brings a considerable fraction of compounds close to or below the hull, suggesting the potential existence of stable MMCH compositions. Our broad screening campaign complements this assessment and provides diffraction evidence consistent with the formation of \ch{Sn2InS2Br3}. We further identify five decomposition reactions through which all MMCHs decompose, with pathway \textit{via} 4\,\ch{M(II)2M(III)Ch2X3} $\leftrightarrow$ 2\,\ch{M(III)_2Ch_3} + 6\,\ch{M(II)X_2} + 2\,\ch{M(II)Ch}, being the most common. Our experiments reproduce the predicted tendency toward binary decomposition products, especially for \ch{Cl}- and \ch{Br}- MMCHs. Moreover, they reveal that \ch{Sn2SbS2I3} forms in high phase purity, \ch{Sn2InS2Br3} is tentatively identified in a multiphase film, and \ch{Sn2SbS2Cl3} is at best transient. Also, our broad screening campaign confirms that these trends are systematic rather than artefacts of specific synthesis conditions. Thus, proximity to the convex hull is most useful as a prioritisation criterion when considered together with the identity and chemical accessibility of the surrounding competing phases. Overall, our work shows that the experimental accessibility of MMCHs is governed not only by the energy of the target phase itself, but critically by the competing low-energy phases and the processing conditions used to access the target. 

\section*{Supporting Information}
The authors have cited additional references within the Supporting Information.\cite{kresse1993, kresse1994, kresse1996, kresse1996a, kresse1999, blochl1994, jain2013, perdew2008, perdew2009, langreth1980, breiman2001, james2013, scikitlearn, lundberg2017, lundberg2018,lundberg2020, degen2014, fullprof, rodriguez1990, bijelic2019, hill1987, 2dprojectiongit}

\section*{Author Contributions}
P.H. and P.R. conceived the project. M.A.L.M. performed the DFT calculations and computed the compositional phase diagrams. D.R.F. and E.S. synthesised the six MMCH compounds. G.K.G., J.La., J.Li, D.R.F., P.V., and P.R. contributed significantly to the scientific discussion of the results. All authors commented on the manuscript, which was drafted by P.H.

\section*{Acknowledgements}
The authors thank Milica Todorovi\'{c} and Armi Tiihonen for discussions. This study was supported by the Research Council of Finland through Project No. 334532, 347772 and the Photonics Research and Innovation (PREIN) Flagship Programme (Grant No. 346511), the European Union's Horizon 2020 research and innovation programme under the Marie Sk\l{}odowska-Curie grant agreement No. 101152684 and the ERC-Consolidator grant agreement number No. 866018, the National Natural Science Foundation of China (Grant No. 22473088), the Natural Science Foundation of Shaanxi Province of China (Grant No. 2023-YBGY-447), abd the Science and Innovation Ministry of Spain/FEDER projects No. PID2023-148976OB-C41 (CURIO-CITY). This work is also part of Maria de Maeztu Units of Excellence Programme CEX2023-001300-M/funded by MICIU/AEI/ 10.13039/501100011033. P.H., J.La., J.Li, M.A.L.M. and P.R. further acknowledge CSC-IT Center for Science, Finland, the Aalto Science-IT project, Xi'an Jiaotong University's HPC platform and the Computing Center in Xi’an of China for generous computational resources.  E.S. is grateful to ICREA Academia program. D.R.F. acknowledges support by Generalitat de Catalunya under grant FI SDUR 2023 (Reference No. BDNS 699313 and DOGC No. 8929 – 02.06.2023).

\section*{Conflict of Interest}

The authors declare no conﬂict of interest.

\section*{Data Availability Statement}
The data that support the findings of this article are openly available: the computed phase diagrams for all 54 MMCH compounds,\cite{mmchsphasediagrams} the code to generate the two-dimensional surface projections,\cite{2dprojectiongit} as well as the measured XRD patterns and the corresponding Rietveld refinements for \ch{Sn2SbS2I3}, \ch{Sn2InS2Br3}, \ch{Sn2InSe2Br3}, \ch{Sn2SbS2Cl3}, \ch{Sn2SbSe2Cl3} \&{} \ch{Sn2InSe2Cl3}.\cite{mmchsxrds}

\bibliography{lib_manuscript.bib}

\end{document}


\newpage

\pdfbookmark[section]{Computational details}{s1}
\section{S1: Computational details}\label{sec:si:s1}

\pdfbookmark[subsection]{Density functional theory}{s1a}
\subsection{I: Density functional theory}\label{sec:si:s1:dft}

Periodic density functional theory calculations were performed using the  Vienna ab initio simulation package (VASP) version 5.2.\cite{kresse1993, kresse1994, kresse1996, kresse1996a} The projector augmented-wave (PAW)\cite{kresse1999, blochl1994} method was employed to describe the electron–ion interactions. All parameters, including the pseudopotentials, were chosen to ensure compatibility with the data available in the Materials Project database.\cite{jain2013} The Perdew-Burke-Ernzerhof (PBE)\cite{perdew2008, perdew2009} exchange-correlation functional was used within the generalized gradient approximation (GGA).\cite{langreth1980} Brillouin-zone sampling was performed using uniform $\Gamma$-centered \textit{k}-point grids with a density of 1000 \textit{k}-points per reciprocal atom. A plane-wave kinetic energy cutoff of \SI{520}{\electronvolt} was employed for all MMCH calculations. The convergence threshold for the ionic relaxation was set to less than \SI{5e-2}{\electronvolt\per\angstrom}.

In order to construct the compositional phase diagrams and their associated convex hulls at $T = \SI{0}{\kelvin}$ and $P = \SI{0}{\atm}$, we calculated the formation energy with respect to elemental decomposition for each MMCH phase as follows:

\begin{equation}
    E_\text{form} = E_\text{total} - \sum_i^N n_i \mu_i
\end{equation}%
with $E_\text{total}$ denotes the DFT-calculated total
energy of the individual MMCH compound, $N$ being the number of elements in the system, $n_i$ the stoichiometric coefficient of component $i$, and $\mu_i$ the corresponding chemical potential.

\pdfbookmark[subsection]{Random forest regression modeling and Shapley additive explanations analysis}{s1b}
\subsection{II: Random forest regression modeling and Shapley explanations analysis}\label{sec:si:s1:rf_shap}

In our previous study (see Ref.\,\citenum{henkel2025}), we investigated the influence of the four atomic sites (\ch{M(II)}, \ch{M(III)}, \ch{Ch}, and \ch{X}) in MMCH compounds and the effect of substitution elements at these sites. Their impact on formation energies, band gaps, and effective carrier masses was analysed using random forest (RF) regression models and Shapley additive explanations (SHAP) analysis. Here, we extend this approach to computed hull energies. The aim of this approach is to characterise the influence of the individual atomic sites, as well as the local contribution of the constituent elements, on the hull energies and, consequently, on the thermodynamic stability.

\begin{figure}[h!]
	\centering
	\includegraphics[width=0.95\textwidth]{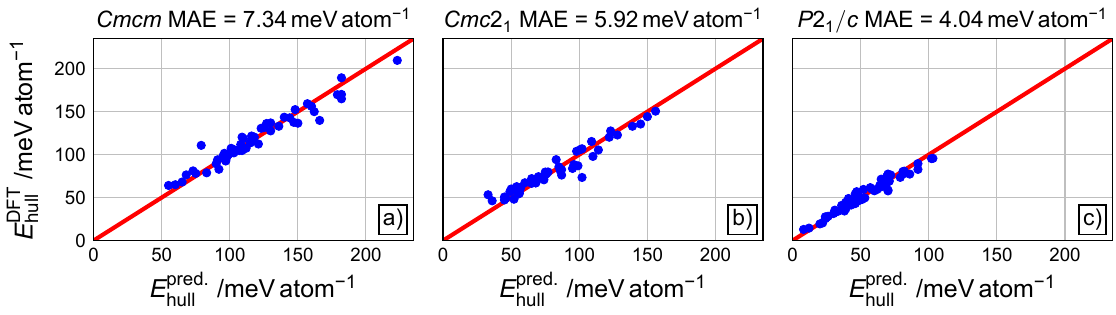}
	\caption{Machine learning (RF regression) predicted hull energies compared to DFT computed hull energies within a) \cmcm, b) \cmcii as well as \piic phase.}
	\label{fig:si:rf_error}
\end{figure}

RF regression, a non-linear ensemble learning method based on decision trees,\cite{breiman2001} predicts regression targets by averaging the outputs of individual trees, thereby improving predictive accuracy and reducing overfitting.\cite{james2013} Atomic numbers were used as compositional descriptors (e.g., \ch{Sn2InS2Br3} $\rightarrow$ [50, 49, 16, 35]) to predict hull energies ($E^\text{pred.}_\text{hull}$). The dataset was first split into a $\SI{75}{\percent}/\SI{25}{\percent}$ training and test set. For each space group, separate RF models were trained using grid search with 10-fold cross-validation (\textit{GridSearchCV}) implemented in \textit{scikit-learn}.\cite{scikitlearn} The best models were selected based on the mean cross-validated root mean square error on the training set, see Figure\,\ref{fig:si:rf_error}. Feature importance analysis of the trained RF models quantifies the contribution of each atomic site to the hull energies, with the corresponding results shown in panel a) of Figures\,\ref{fig:si:shap_cmcm} ($Cmcm$), \ref{fig:si:shap_cmc21} ($Cmc2_1$) and \ref{fig:si:shap_p21c} ($P2_1/c$).

Furthermore, to analyse the impact of elements on the hull energies within the MMCH material space, the trained RF models were further interpreted using SHAP.\cite{lundberg2017} SHAP is a game theory-based feature attribution method that quantifies the contribution of each input feature to the predicted target value. In this work, we employ the \textit{TreeSHAP} approach, which is specifically designed for tree-based models,\cite{lundberg2018,lundberg2020} to assess the influence of the features (atomic numbers of elements within the four atom sites of MMCH compounds) on the predicted hull energies. The SHAP analysis is performed relative to the model base value, defined as the mean prediction of each RF model. The corresponding results are shown in panel b) of Figures\,\ref{fig:si:shap_cmcm} ($Cmcm$), \ref{fig:si:shap_cmc21} ($Cmc2_1$) and \ref{fig:si:shap_p21c} ($P2_1/c$).

\clearpage

\pdfbookmark[section]{Experimental details}{s2}
\section{S2: Experimental details}\label{sec:si:s2}

\pdfbookmark[subsection]{Mixed-metal chalcohalides synthesis}{s2a}
\subsection{I: Mixed-metal chalcohalides synthesis}\label{sec:si:s2:mmch_synthesis}

\hspace{.5cm}{\textbf{a) Precursor information}}\\[-.25cm]

\noindent Tin(II)chloride anhydrous powder ($>$\SI{99.99}{\percent}), thiourea (\ch{SC(NH2)2}, $>$\SI{99}{\percent}), \textit{N},\textit{N}-Dimethylformamide, antimony(III)chloride ($>$\SI{99}{\percent}), indium(III)chloride (\SI{99.999}{\percent}), tin(II)bromide ($>$\SI{99}{\percent}) and propylamine ($>$\SI{99}{\percent}) were purchased from Sigma-Aldrich. Tin powder (\SI{99.995}{\percent}), tin(II)selenide (\SI{99.999}{\percent}), indium(III)acetate (\SI{99.99}{\percent}), antimony(III)acetate (\SI{99.99}{\percent}), selenourea (\ch{SeC(NH2)2}, \SI{98}{\percent}), tin(II)iodide (\SI{99.999}{\percent}), 1,2-ethanedithiol ($+$\SI{98}{\percent}) and titanium(diisopropoxide)bis(2,4-pentanedionate) (\SI{75}{\percent} in ethanol) were purchased from ThermoScientific.

\vspace{.5cm}

\hspace{.5cm}{\textbf{b) Substrate preparation}}\\[-.25cm]

\noindent Florine-doped tin oxide (FTO) substrates and molybdenum substrates were purchased from Sigma-Aldirch and Suzhou ShangYang Solar Technology, respectively. Titanium dioxide films were prepared \textit{via} spray pyrolysis of a \SI{5}{\percent} ethanol solution of Titanium(diisopropoxide)bis(2,4-pentanedionate) over FTO substrate, followed by annealing at \SI{500}{\celsius} for \SI{20}{\min}.

\vspace{.5cm}

\hspace{.5cm}{\textbf{c) Film synthesis}}\\[-.25cm]

\noindent Film synthesis has been done through molecular ink deposition. Precursor salts were dissolved in \textit{N},\textit{N}-Dimethylformamide (DMF) at different concentrations and ratios to study different growth conditions. The \ch{M(III)} precursor concentration is fixed at \SI{0.2}{\mol\per\l}, and the precursors are sequentially dissolved in the following order: \ch{M(III)}\;salt\;$\rightarrow$\;\ch{Ch}\;salt\;$\rightarrow$\;\ch{M(II)}\;salt. After dissolution, \SI{50}{\micro\l} of the molecular ink is transferred onto the selected substrate using a micropipette. The substrates are pre-cleaned to ensure proper adhesion. The precursor is then spin-coated at \SIrange[range-phrase = {--}, range-units = single]{1000}{2000}{\rpm} for \SI{30}{\sec}, after which the sample is transferred to a hotplate for soft annealing. A pre-annealing step is performed at \SI{150}{\celsius} for \SI{3}{\min} to gently remove the solvent. Subsequently, the temperature is increased to the crystallisation temperature of the material, \SIrange[range-phrase = {--}, range-units = single]{200}{350}{\celsius}, and maintained for \SI{5}{\min}. All the preparation was carried out in an inert atmosphere (using a glovebox system) with controlled oxygen and water levels ($<$\SI{1}{} and $<$\SI{30}{\ppm}, respectively).

For the \ch{Sn2SeIn2Cl3} and \ch{Sn2SeSb2Cl3} compounds a thiol-amine solvent system was used as an explorative alternative. The solution preparation is performed under inert atmosphere, and it consisted of mixing propylamine (PA) with 1,2-ethanedithiol (EDT) in a 3:1 ratio. Under continuous stirring at \SI{35}{\celsius}, \ch{SnSe} and \ch{SbCl3} or \ch{InCl3} powder were dissolved in the EDT+PA mixture. The resulting solution was deposited onto the chosen substrate following the previously described procedure.

\newpage

\hspace*{-.75cm}
\begin{minipage}{1.0\textwidth}
\centering
\captionof{table}{Summary of the experimental synthesis strategies in DMF performed for \ch{Sn2SbS2I3}, \ch{Sn2InS2Br3}, \ch{Sn2InSe2Br3}, \ch{Sn2SbS2Cl3}, \ch{Sn2SbSe2Cl3} and \ch{Sn2InSe2Cl3}. Phase identification was carried out with \textsc{X'Pert HighScore Plus} (version 4.8) in combination with reference patterns from the PDF-2 database (2018 release).\cite{degen2014} Samples highlighted in cyan were selected for subsequent Rietveld refinement, see Tables \ref{tab:si:rietfeld_refdata:sn2sbs2i3} - \ref{tab:si:rietfeld_refdata:sn2inse2cl3}.}
\label{tab:si:mmch_exp_samples}
\scalebox{.79}{
\begin{tabular}{C{2.25cm}||C{1.75cm}|C{6.1cm}|C{2.1cm}|C{2.35cm}|C{3.35cm}} 
\textbf{MMCH} & \textbf{Sample number} & \textbf{Precursor strategy} & \textbf{Substrate} & \textbf{Annealing temp.} /\unit{\celsius} & \textbf{Identified phases} \\\hline\hline

\vspace*{-.25cm}\scalebox{1.05}{\ch{Sn2SbS2I3}} & \cellcolor{cyan!50} 1 & \cellcolor{cyan!50} standard \ch{TU} route (\ch{SnI2} + \num{0.7}\,\ch{SbCl3} + \num{1.8}\,\ch{TU}) & \cellcolor{cyan!50} FTO/\ch{TiO2}/ mp-\ch{TiO2} & \cellcolor{cyan!50} \num{300} & \cellcolor{cyan!50} \ch{Sn2SbS2I3}, \ch{SnO2}, \ch{SnS} \\\hline\hline

\multirow{3}{*}{\vspace*{-1.5cm}\scalebox{1.05}{\ch{Sn2InS2Br3}}\footnote{Identification of \ch{Sn2InS2Br3} \textit{via} peak matching was not possible, because there is not reference
pattern included in the PDF-2 database (2018 release).\cite{degen2014}}}
	& 2 		& standard \ch{TU} route (\ch{SnBr2} + \num{0.7}\,\ch{InCl3} + \num{1.8}\,\ch{TU}) 					& FTO 							& \num{250} 		& \ch{SnS}, \ch{SnO}, \ch{InCl} \\
	& \cellcolor{cyan!50} 3 & \cellcolor{cyan!50} standard \ch{TU} route (\ch{SnBr2} + \num{0.7}\,\ch{InCl3} + \num{1.8}\,\ch{TU}) & \cellcolor{cyan!50} FTO & \cellcolor{cyan!50} \num{300} & \cellcolor{cyan!50} \ch{SnS}, \ch{In2S3}, \ch{InCl} \\
    & 4 & standard \ch{TU} route (\ch{SnBr2} + \num{0.7}\,\ch{InCl3} + \num{1.8}\,\ch{TU}) & FTO & \num{300} & \ch{SnS2}, \ch{In2S3}, \ch{SnO2}, \ch{InCl} \\\hline\hline

\multirow{9}{*}{\vspace*{-5cm}\scalebox{1.05}{\ch{Sn2InSe2Br3}}}
	& 5 		& standard \ch{SeU} route (\ch{SnBr2} + \num{0.7}\,\ch{InCl3} + \num{1.8}\,\ch{SeU})						& \ch{Mo}						& \num{300} 		& \ch{SnSe}, \ch{Mo}\\
	& \cellcolor{cyan!50} 6 		& \cellcolor{cyan!50} standard \ch{SeU} route (\ch{SnBr2} + \num{0.7}\,\ch{InCl3} + \num{1.8}\,\ch{SeU})						& \cellcolor{cyan!50} \ch{Mo} 						& \cellcolor{cyan!50} \num{350} 		& \cellcolor{cyan!50} \ch{SnSe}, \ch{SnSe2}, \ch{In2Se3}, \ch{Mo} \\\cline{2-6}
 	& 7		& standard \ch{SeU} route (\ch{SnBr2} + \num{0.7}\,\ch{InCl3} + \num{1.8}\,\ch{SeU})						& FTO/\ch{TiO2}/ mp-\ch{TiO2}	& \num{300} 		& \ch{SnO2}, \ch{SnSe}, \ch{InSeBr}, \ch{TiO2} \\\cline{2-6}
 	& 8		& Acetate-substituted \ch{In} precursor route (\num{2}\,\ch{SnBr2} + \ch{In(OAc)3} + \num{2}\,\ch{SeU})	& \ch{Mo} 						& \num{200} 		& Amorphous \\
 	& 9		& Acetate-substituted \ch{In} precursor route (\num{2}\,\ch{SnBr2} + \ch{In(OAc)3} + \num{2}\,\ch{SeU})	& \ch{Mo} 						& \num{300} 		& Amorphous \\\cline{2-6}
 	& 10	& \ch{In}-lean acetate route (\num{2}\,\ch{SnBr2} + \num{0.5}\,\ch{In(OAc)3} + \num{1.5}\,\ch{SeU})		& \ch{Mo} 						& \num{200} 		& Amorphous \\
 	& 11		& \ch{In}-lean acetate route (\num{2}\,\ch{SnBr2} + \num{0.5}\,\ch{In(OAc)3} + \num{1.5}\,\ch{SeU})		& \ch{Mo} 						& \num{300}		& \ch{SnSe}, \ch{Mo} \\\cline{2-6}
 	& 12 	& \ch{Br}-rich / \ch{Sn}-rich route (\num{3}\,\ch{SnBr2} + \num{0.5}\,\ch{InBr3} + \num{2}\,\ch{SeU}) 		& \ch{Mo} 						& \num{200} 		& Amorphous \\
 	& 13		& \ch{Br}-rich / \ch{Sn}-rich route (\num{3}\,\ch{SnBr2} + \num{0.5}\,\ch{InBr3} + \num{2}\,\ch{SeU}) 		& \ch{Mo} 						& \num{300} 		& \ch{SnSe}, \ch{SnSe2}, \ch{Mo} \\\hline\hline

\multirow{2}{*}{\vspace*{-1cm}\scalebox{1.05}{\ch{Sn2SbS2Cl3}}}
	& 14		& standard \ch{TU} route (\ch{SnCl2} + \num{0.7}\,\ch{SbCl3} + \num{1.8}\,\ch{TU}) 					& FTO 							& \num{250} 		& \ch{SnO2}, \ch{SnS}, \ch{Sn2S3}, \ch{Sb2O4} \\
	& \cellcolor{cyan!50} 15 & \cellcolor{cyan!50} standard \ch{TU} route (\ch{SnCl2} + \num{0.7}\,\ch{SbCl3} + \num{1.8}\,\ch{TU}) & \cellcolor{cyan!50} FTO & \cellcolor{cyan!50} \num{300} & \cellcolor{cyan!50} \ch{SnO2}, \ch{SnS}, \ch{Sn2S3}, \ch{Sb2S3} \\\hline\hline

\multirow{2}{*}{\vspace*{-1.0cm}\scalebox{1.05}{\ch{Sn2SbSe2Cl3}}}
	& 16 		& standard \ch{SeU} route (\ch{SnCl2} + \num{0.7}\,\ch{SbCl3} + \num{1.8}\,\ch{SeU})					& FTO							& \num{300} 		& \ch{SnSe}, \ch{Mo} \\\cline{2-6}
 	& \cellcolor{cyan!50} 17 & \cellcolor{cyan!50} \ch{Sn}-rich \ch{SeU} route (\num{1.2}\,\ch{SnCl2} + \num{0.7}\,\ch{SbCl3} + \num{1.8}\,\ch{SeU}) & \cellcolor{cyan!50} FTO	& \cellcolor{cyan!50} \num{300} & \cellcolor{cyan!50} \ch{SnSe}, \ch{Sb2Se3} \\\cline{2-6}

\end{tabular}}
\end{minipage}


\hspace*{-.75cm}
\begin{minipage}{1.0\textwidth}
\centering
\captionof*{table}{Continuation from Table\,\ref{tab:si:mmch_exp_samples}: Summary of the experimental synthesis strategies in DMF performed for \ch{Sn2SbS2I3}, \ch{Sn2InS2Br3}, \ch{Sn2InSe2Br3}, \ch{Sn2SbS2Cl3}, \ch{Sn2SbSe2Cl3} and \ch{Sn2InSe2Cl3}. Phase identification was carried out with \textsc{X'Pert HighScore Plus} (version 4.8) in combination with reference patterns from the PDF-2 database (2018 release).\cite{degen2014} Samples highlighted in cyan were selected for subsequent Rietveld refinement, see Tables \ref{tab:si:rietfeld_refdata:sn2sbs2i3} - \ref{tab:si:rietfeld_refdata:sn2inse2cl3}.}
\scalebox{.79}{
\begin{tabular}{C{2.25cm}||C{1.75cm}|C{6.1cm}|C{2.1cm}|C{2.35cm}|C{3.35cm}} 
\textbf{MMCH} & \textbf{Sample number} & \textbf{Precursor strategy} & \textbf{Substrate} & \textbf{Annealing temp.} /\unit{\celsius} & \textbf{Identified phases} \\\hline\hline


\multirow{12}{*}{\vspace*{-5.5cm}\scalebox{1.05}{\ch{Sn2SbSe2Cl3}}}
 	& 18		& More \ch{Sn}-rich \ch{SeU} route (\num{1.7}\,\ch{SnCl2} + \num{0.7}\,\ch{SbCl3} + \num{1.8}\,\ch{SeU})				& FTO							& \num{300} 		& \ch{SnSe}, \ch{Sb2Se3} \\\cline{2-6}
 	& 19		& Acetate-substituted \ch{Sn} precursor route (\ch{Sn(OAc)2} + \num{0.7}\,\ch{SbCl3} + \num{1.8}\,\ch{SeU})& \ch{Mo}						& \num{200} 		& Amorphous \\
 	& 20		& Acetate-substituted \ch{Sn} precursor route (\ch{Sn(OAc)2} + \num{0.7}\,\ch{SbCl3} + \num{1.8}\,\ch{SeU})& \ch{Mo} 						& \num{300} 		& \ch{SnO2}, \ch{Sb2O3}, \ch{Sb2Se3} \\\cline{2-6}
 	& 21		& Two-vial route with metallic \ch{Sn} additive															& \ch{Mo} 						& \num{200} 		& Amorphous \\
 	& 22		& Two-vial route with metallic \ch{Sn} additive															& \ch{Mo} 						& \num{250}		& Amorphous \\\cline{2-6}
 	& 23 	& Two-vial route without metallic \ch{Sn}																& \ch{Mo} 						& \num{200} 		& Amorphous \\
 	& 24		& Two-vial route without metallic \ch{Sn}																& \ch{Mo} 						& \num{250} 		& \ch{Sb2Se3} \\\cline{2-6}
 	& 25		& direct \ch{SnSe} + \ch{SbCl3} route		 															& FTO	 						& \num{150} 		& \ch{SnO2}, \ch{SnSe}, \ch{Sb2Se3} \\
 	& 26		& direct \ch{SnSe} + \ch{SbCl3} route		 															& FTO	 						& \num{200} 		& \ch{SnO2}, \ch{SnSe}, \ch{Sb2O3} \\
 	& 27		& direct \ch{SnSe} + \ch{SbCl3} route		 															& FTO							& \num{250} 		& \ch{SnO2}, \ch{SnSe} \\\cline{2-6}
 	& 28		& direct \ch{SnSe} + \ch{SbCl3} route\footnote{Synthesis has been performed in amine-thiol solvent (PA:EDT) instead of DMF.} & FTO 					& \num{300} 		& \ch{SnO2}, \ch{Sb2Se3} \\
 	& 29		& standard \ch{SeU} route (\ch{SnCl2} + \num{0.7}\,\ch{InCl3} + \num{1.8}\,\ch{SeU}) 						& FTO/\ch{TiO2}/ mp-\ch{TiO2}	& \num{300} 		& \ch{SnO2}, \ch{SnSe2}, \ch{Sb2Se3}, \ch{TiO2} \\\hline\hline


\multirow{7}{*}{\vspace*{-4cm}\scalebox{1.05}{\ch{Sn2InSe2Cl3}}}
	& 30		& direct \ch{SnSe} + \ch{InCl3} route	 															& FTO 							& \num{150} 		& \ch{SnO2}, \ch{In2O3}, \ch{InSe}, \ch{Se} \\
	& 31 	& direct \ch{SnSe} + \ch{InCl3} route	 															& FTO 							& \num{200} 		& \ch{SnO2}, \ch{In2O3}, \ch{Se}, \ch{Sn} \\
 	& 32		& direct \ch{SnSe} + \ch{InCl3} route	 															& FTO 							& \num{250} 		& \ch{SnO2}, \ch{SnSe}, \ch{In2S3}, \ch{Se}, \ch{Sn} \\\cline{2-6}
 	& 33		& standard \ch{SeU} route (\ch{SnCl2} + \num{0.7}\,\ch{InCl3} + \num{1.8}\,\ch{SeU}) 					& \ch{Mo} 						& \num{250} 		& \ch{SnSe}, \ch{SnSe2}, \ch{MoO3}\\
 	& 34		& standard \ch{SeU} route (\ch{SnCl2} + \num{0.7}\,\ch{InCl3} + \num{1.8}\,\ch{SeU}) 					& \ch{Mo} 						& \num{300} 		& \ch{SnSe}, \ch{SnSe2}, \ch{SnO}, \ch{SnO2}\\
 	&\cellcolor{cyan!50} 35 & \cellcolor{cyan!50} standard \ch{SeU} route (\ch{SnCl2} + \num{0.7}\,\ch{InCl3} + \num{1.8}\,\ch{SeU}) & \cellcolor{cyan!50} \ch{Mo} & \cellcolor{cyan!50} \num{350} & \cellcolor{cyan!50} \ch{SnSe}, \ch{SnSe2}, \ch{In2Se3}, \ch{Mo}\\\cline{2-6}
 	& 36		& standard \ch{SeU} route (\ch{SnCl2} + \num{0.7}\,\ch{InCl3} + \num{1.8}\,\ch{SeU}) 					& FTO/\ch{TiO2}/ mp-\ch{TiO2} 	& \num{350} 		& \ch{SnSe}, \ch{SnSe2}, \ch{SnO2}, \ch{TiSe2}, \ch{TiO2}\\\hline\hline
\end{tabular}}
\end{minipage}

\clearpage

\pdfbookmark[subsection]{X-ray diffraction measurements}{s2b}
\subsection{II: X-ray diffraction measurements}\label{sec:si:s2:xrd}

The phase composition of the synthesised MMCH compounds was determined by X-ray diffraction (XRD). Measurements were carried out on a Bruker D8 Advance diffractometer operating in Bragg-Brentano geometry at \SI{40}{\kilo\volt} and \SI{40}{\milli\ampere}, using CuK$\alpha_1$ radiation ($\lambda = \SI{1.54187}{\angstrom}$). The diffraction patterns were analysed using the \textsc{X'Pert HighScore Plus} (Version 4.8)\cite{degen2014} software package, and phase identification was performed based on reference patterns from the PDF-2 database (2018 release).

\pdfbookmark[subsection]{Rietveld refinement}{s2c}
\subsection{III: Rietveld refinement}\label{sec:si:s2:rietfeld}

\begin{figure}[h!]
	\centering
	\includegraphics[width=0.7375\textwidth]{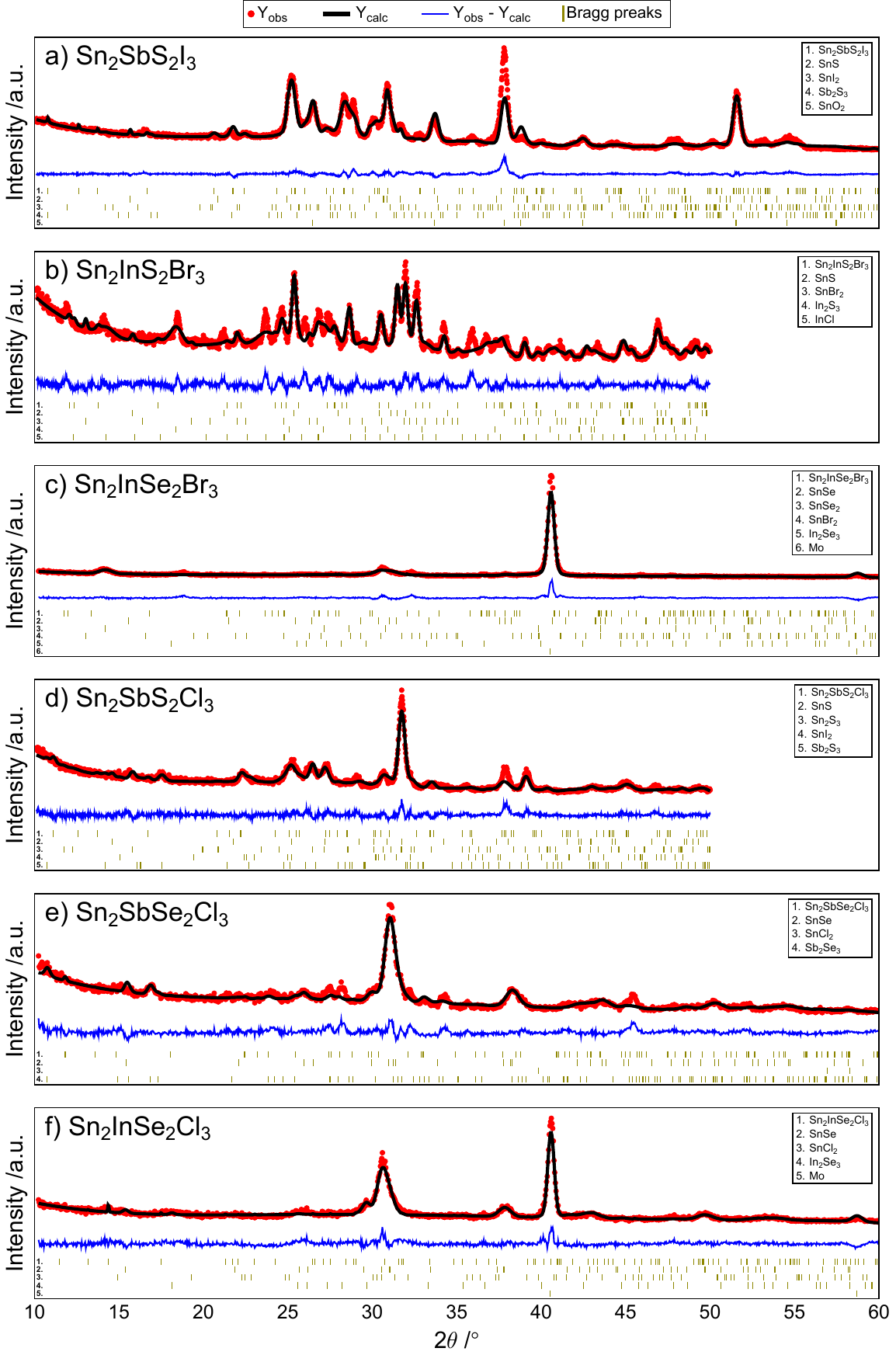}
	\caption{Rietveld refinement of the measured XRD pattern for a) \ch{Sn2SbS2I3}, b) \ch{Sn2InS2Br3}, c) \ch{Sn2InSe2Br3}, d) \ch{Sn2SbS2Cl3}, e) \ch{Sn2SbSe2Cl3}, and f) \ch{Sn2InSe2Cl3}.}
	\label{fig:si:rietfeld}
\end{figure}

For each MMCH compound, one representative X-ray diffraction pattern was selected for refinement based on the prior performed phase identification, see Section \hyperref[sec:si:s8]{S8}. Rietveld refinements were then carried out using the \textsc{Fullprof} software package (Version 5.20, December 2023),\cite{fullprof} see Figure \ref{fig:si:rietfeld} for a) \ch{Sn2SbS2I3}, b) \ch{Sn2InS2Br3}, c) \ch{Sn2InSe2Br3}, d) \ch{Sn2SbS2Cl3}, e) \ch{Sn2SbSe2Cl3} and f) \ch{Sn2InSe2Cl3}. The modified Thompson-Cox-Hastings pseudo-Voigt function was employed as the profile function,\cite{rodriguez1990} as it provides a reliable description of peak shapes and enables the separation of size and micro-strain contributions to peak broadening.\cite{bijelic2019} A sixth-order polynomial was employed to model the background. The phase composition was determined following the procedure of Hill and Howard,\cite{hill1987} see Table \ref{tab:si:phase_comp} for the actual phase compositions (including substrate, oxides and reaction by-products contributions if present). The refinements converged with satisfactory agreement factors, see Tables \ref{tab:si:rietfeld_refdata:sn2sbs2i3} to \ref{tab:si:rietfeld_refdata:sn2inse2cl3}.

\clearpage

\pdfbookmark[section]{Computed phase diagrams for \protect\ch{M(II)2M(III)Ch2X3} compounds}{s3}
\section{S3: Computed phase diagrams for \protect\ch{M(II)2M(III)Ch2X3} compounds}\label{sec:si:s3}

\begin{figure}[h!]
	\centering
	\includegraphics[width=1.0\textwidth]{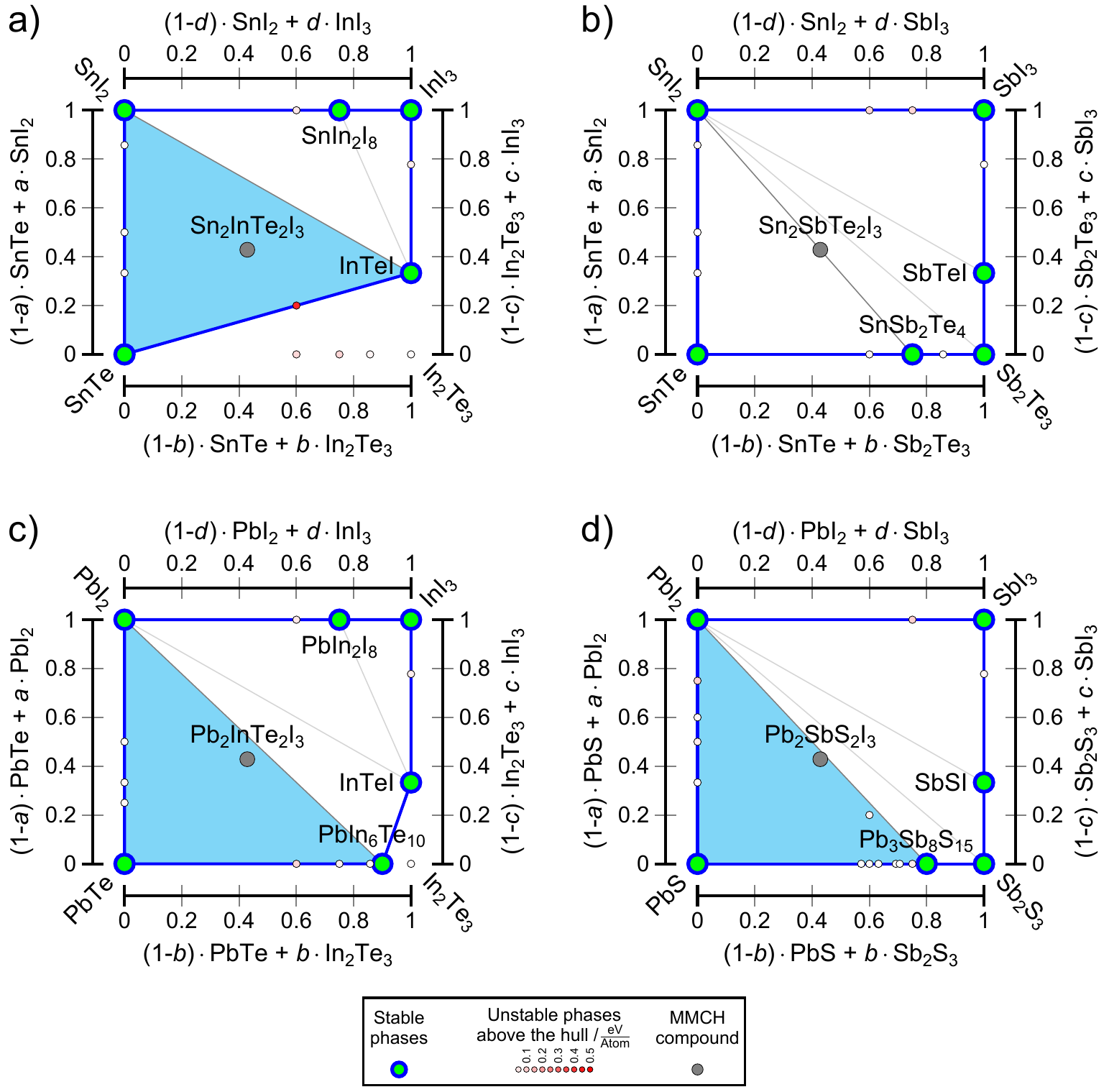}
	\caption{Two-dimensional surface projection of the four-dimensional phase diagram for MMCH compounds decomposing according to reactions\,(2), (3), (4), and (5). The two-dimensional surface is spanned by \ch{M(II)X2}, \ch{M(III)X3}, \ch{M(II)Ch}, and \ch{M(III)2Ch3}, analogous to the procedure shown in Figure\,1. a) Reaction (2), exemplified by \ch{Sn2InTe2I3}, b) Reaction (3), exemplified by \ch{Sn2SbTe2I3}, c) Reaction (4), exemplified by \ch{Pb2InTe2I3}, and d) Reaction (5), exemplified by \ch{Pb2SbS2I3}. Color code: Cyan-coloured triangle: convex hull region (MMCH compound decomposes into three products); Green: stable phases; Shades of red: unstable phases ($E_\text{hull} \leq \SI{0.5}{\electronvolt\per\at}$) according to there $E_\text{hull}$ value; Gray: MMCH compounds.}
	\label{fig:si:2d_phasediagram_projection}
\end{figure}

To simplify the visualisation of the decomposition products of MMCH compounds, we perform a two-dimensional surface projection of the four-dimensional phase diagram. All MMCHs are located either on a convex hull surface (cyan-colored triangle in Figure\,1 and SI Figure\,\ref{fig:si:2d_phasediagram_projection} a), c) \&{} d) in the case MMCHs decompose into three phases) or along a convex hull line (gray-colored line in SI Figure\,\ref{fig:si:2d_phasediagram_projection} b) in the case MMCH compounds decompose into two phases). Therefore, it is sufficient to consider only the two-dimensional surface spanned by \ch{M(II)X2}, \ch{M(III)X3}, \ch{M(II)Ch}, and \ch{M(III)2Ch3}. In this way, all relevant stable decomposition products, as well as other metastable or unstable phases, can be clearly represented for each MMCH compound. The corresponding implementation to generate these surface projections is provided in Ref.\,\citenum{2dprojectiongit}. Note, in the cases of decomposition reactions\,(2) and (4), it is possible that the phase \ch{M(III)2Ch3} is not stable, but is nevertheless used as a corner point of the two-dimensional surface projection, see Figure\,\ref{fig:si:2d_phasediagram_projection} a) \&{} c).\\

\begin{figure}[h!]
	\centering
	\includegraphics[width=1.0\textwidth]{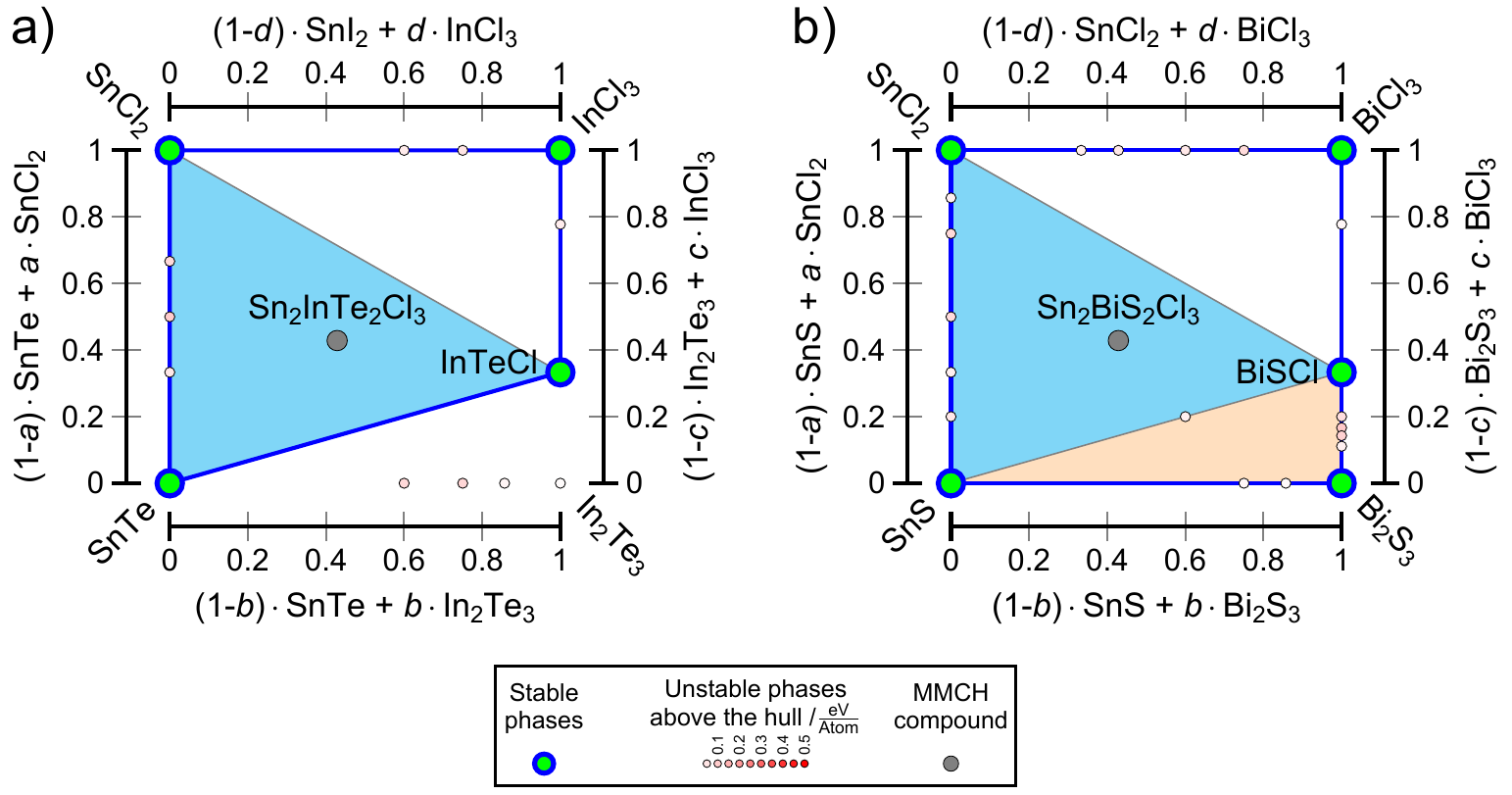}
	\caption{Comparison of the two ways by which MMChs decompose after reaction\,(2): (a) when \ch{M(III)2Ch3} is unstable, exemplified by \ch{Sn2InTe2Cl3}, and (b) when \ch{M(III)2Ch3} is stable, exemplified by \ch{Sn2BiS2Cl3}. In (b), the stability of \ch{M(III)2Ch3} results in a stable \ch{SnCh} - \ch{M(III)2Ch3} - \ch{BiChX} hull surface (light orange triangle). The two-dimensional surface is spanned by \ch{M(II)X2}, \ch{M(III)X3}, \ch{M(II)Ch}, and \ch{M(III)2Ch3}, analogous to the procedure shown in Figure\,1. Color code: Cyan-coloured triangle: convex hull region (MMCH compound decomposes into three products); Light orange-coloured triangle: additional convex hull region (relevant for the decomposition of MMCHs according to reaction\,(2) in the case  \ch{M(III)2Ch3} is stable); Green: stable phases; Shades of red: unstable phases ($E_\text{hull} \leq \SI{0.5}{\electronvolt\per\at}$) according to there $E_\text{hull}$ value; Gray: MMCH compounds.}
	\label{fig:si:2d_pd_proj_reaction2_vgl}
\end{figure}

As described in the main text, the decomposition of MMCH compounds can be summarised by five different reactions, see also Figure\,3. The corresponding two-dimensional surface projection for reaction 1 is shown in Figure\,1 exemplified by \ch{Sn2InS2Br3}. Furthermore, Figure\,\ref{fig:si:2d_phasediagram_projection} shows the additional two-dimensional surface projections for
a) reaction (2), exemplified by \ch{Sn2InTe2I3}, b) reaction (3), exemplified by \ch{Sn2SbTe2I3}, c) reaction (4), exemplified by \ch{Pb2InTe2I3}, and d) reaction (5), exemplified by \ch{Pb2SbS2I3}. In addition, Figure\,\ref{fig:si:2d_pd_proj_reaction2_vgl} illustrates the two-dimensional surface projections for reaction (2) as a function of the stability of \ch{M(III)2Ch3}.

\clearpage

\clearpage

\pdfbookmark[section]{Tabulated hull energies and decomposition reactions for \protect\ch{M(II)2M(III)Ch2X3} compounds}{s4}
\section{S4: Tabulated hull energies and decomposition reactions for \protect\ch{M(II)2M(III)Ch2X3} compounds}\label{sec:si:s4}

\begin{table}[h!]
\centering
\caption{DFT (PBE level of theory) calculated hull energies ($E_\text{hull}$), along with the corresponding decomposition reactions and reaction types (1) to (5), as shown in Figure\,3 of the main text, for all \protect\ch{M(II)2M(III)Ch2X3} compounds. In parentheses: Hull energies from the Materials Project\cite{materialsproject2013} if available. Colors: Purple: MMCH compounds that have been studied experimentally and/or theoretically in the literature; Cyan: lead-free MMCH compounds that are experimentally investigated in this work.}
\label{tab:si:hulle_decompreac}
\scalebox{.8125}{
\begin{tabular}{C{2.75cm}|C{1.275cm}C{1.275cm}C{1.275cm}||C{1.0cm}|C{9.25cm}}
~\linebreak ~\linebreak \textbf{MMCH} & \textbf{\textit{Cmcm}} \textbf{\textit{E}\textsubscript{hull}} \textbf{/\unit[per-mode=fraction]{\milli\electronvolt\per\at}} & \textbf{\textit{Cmc}2\textsubscript{1}} \textbf{\textit{E}\textsubscript{hull}} \textbf{/\unit[per-mode=fraction]{\milli\electronvolt\per\at}} & \textbf{\textit{P}2\textsubscript{1}\textit{/c}} \textbf{\textit{E}\textsubscript{hull}} \textbf{/\unit[per-mode=fraction]{\milli\electronvolt\per\at}} & ~\linebreak ~\linebreak \textbf{Type} & ~\linebreak ~\linebreak \textbf{Decomposition reaction} \\\hline\hline
\ch{Sn2InS2Cl3} 	& \num{95} 				& \num{46} 				& \num{70} 				& \num{1} 	& 4\,\ch{Sn2InS2Cl3} 	$\leftrightarrow$ 2\,\ch{In2S3} 		+ 6\,\ch{SnCl2}	+ 2\,\ch{SnS} \\
\rowcolor{cyan!50}
\ch{Sn2InS2Br3} 	& \num{95} 				& \num{35} 				& \num{41} 				& \num{1} 	& 4\,\ch{Sn2InS2Br3} 	$\leftrightarrow$ 2\,\ch{In2S3} 		+ 6\,\ch{SnBr2}	+ 2\,\ch{SnS} \\
\ch{Sn2InS2I3} 	    & \num{108} 			& \num{102} 			& \num{70} 				& \num{1} 	& 4\,\ch{Sn2InS2I3} 	$\leftrightarrow$ 2\,\ch{In2S3} 		+ 6\,\ch{SnI2} 	+ 2\,\ch{SnS} \\
\rowcolor{cyan!50}
\ch{Sn2InSe2Cl3}	& \num{112} 			& \num{57} 				& \num{79} 				& \num{1}	& 4\,\ch{Sn2InSe2Cl3} 	$\leftrightarrow$ 2\,\ch{In2Se3} 	+ 6\,\ch{SnCl2} 	+ 2\,\ch{SnSe} \\
\rowcolor{cyan!50}
\ch{Sn2InSe2Br3}	& \num{97} 			    & \num{59} 				& \num{70} 				& \num{1} 	& 4\,\ch{Sn2InSe2Br3} 	$\leftrightarrow$ 2\,\ch{In2Se3} 	+ 6\,\ch{SnBr2} 	+ 2\,\ch{SnSe} \\
\ch{Sn2InSe2I3} 	& \num{103} 			& \num{60} 				& \num{46} 				& \num{1} 	& 4\,\ch{Sn2InSe2I3} 	$\leftrightarrow$ 2\,\ch{In2Se3} 	+ 6\,\ch{SnI2}	+ 2\,\ch{SnSe} \\
\ch{Sn2InTe2Cl3}	& \num{179} 			& \num{78} 				& \num{103} 			& \num{2} 	& 2\,\ch{Sn2InTe2Cl3} 	$\leftrightarrow$ 2\,\ch{InTeCl} 	+ 2\,\ch{SnCl2} + 2\,\ch{SnTe} \\
\ch{Sn2InTe2Br3}	& \num{156} 			& \num{65} 				& \num{91} 				& \num{2} 	& 2\,\ch{Sn2InTe2Br3} 	$\leftrightarrow$ 2\,\ch{InTeBr} 	+ 2\,\ch{SnBr2} + 2\,\ch{SnTe} \\
\ch{Sn2InTe2I3} 	& \num{101} 			& \num{72} 				& \num{103} 			& \num{2} 	& 2\,\ch{Sn2InTe2I3} 	$\leftrightarrow$ 2\,\ch{InTeI} 		+ 2\,\ch{SnI2} 	+ 2\,\ch{SnTe} \\\hline

\rowcolor{cyan!50}
\ch{Sn2SbS2Cl3} 	& \num{99} 				& \num{74} 				& \num{41} 				& \num{1} 	& 4\,\ch{Sn2SbS2Cl3} 	$\leftrightarrow$ 2\,\ch{Sb2S3} 		+ 6\,\ch{SnCl2} 	+ 2\,\ch{SnS} \\
\ch{Sn2SbS2Br3} 	& \num{96} 				& \num{52} 				& \num{8} 				& \num{1} 	& 4\,\ch{Sn2SbS2Br3} 	$\leftrightarrow$ 2\,\ch{Sb2S3} 		+ 6\,\ch{SnBr2} 	+ 2\,\ch{SnS} \\
\rowcolor{purple!50}
\ch{Sn2SbS2I3} 	    & \num{90} (\num{89})	& \num{57} (\num{56})	& \num{12} 				& \num{1} 	& 4\,\ch{Sn2SbS2I3} 		$\leftrightarrow$ 2\,\ch{Sb2S3} 		+ 6\,\ch{SnI2} 	+ 2\,\ch{SnS} \\
\rowcolor{cyan!50}
\ch{Sn2SbSe2Cl3}	& \num{109} 			& \num{95} 				& \num{48} 				& \num{1} 	& 4\,\ch{Sn2SbSe2Cl3} 	$\leftrightarrow$ 2\,\ch{Sb2Se3} 	+ 6\,\ch{SnCl2} + 2\,\ch{SnSe} \\
\ch{Sn2SbSe2Br3}	& \num{75} 				& \num{69} 				& \num{24} 				& \num{1} 	& 4\,\ch{Sn2SbSe2Br3} 	$\leftrightarrow$ 2\,\ch{Sb2Se3} 	+ 6\,\ch{SnBr2} + 2\,\ch{SnSe} \\
\rowcolor{purple!50}
\ch{Sn2SbSe2I3} 	& -/- 					& \num{69} (\num{61})	& \num{25}  				& \num{1} 	& 4\,\ch{Sn2SbSe2I3} 	$\leftrightarrow$ 2\,\ch{Sb2Se3} 	+ 6\,\ch{SnI2} + 2\,\ch{SnSe} \\
\ch{Sn2SbTe2Cl3}	& \num{130} 			& \num{150} 			& \num{72} 				& \num{3} 	& 4\,\ch{Sn2SbTe2Cl3} 	$\leftrightarrow$ 2\,\ch{SnSb2Te4} 	+ 6\,\ch{SnCl2} \\
\ch{Sn2SbTe2Br3}	& -/- 					& \num{109}				& \num{57}  			& \num{3} 	& 4\,\ch{Sn2SbTe2Br3} 	$\leftrightarrow$ 2\,\ch{SnSb2Te4} 	+ 6\,\ch{SnBr2} \\
\ch{Sn2SbTe2I3}	    & \num{162} 			& \num{102} 			& \num{59} 				& \num{3} 	& 4\,\ch{Sn2SbTe2I3} 	$\leftrightarrow$ 2\,\ch{SnSb2Te4} 	+ 6\,\ch{SnI2} \\\hline
\ch{Sn2BiS2Cl3}	& \num{73} 				& \num{50} 				& \num{35} 				& \num{2} 	& 2\,\ch{Sn2BiS2Cl3} 	$\leftrightarrow$ 2\,\ch{BiSCl} 		+ 2\,\ch{SnCl2} 	+ 2\,\ch{SnS} \\
\ch{Sn2BiS2Br3}	& -/- 					& \num{56} 				& \num{20}  				& \num{2} 	& 2\,\ch{Sn2BiS2Br3} 	$\leftrightarrow$ 2\,\ch{BiSBr} 		+ 2\,\ch{SnBr2} 	+ 2\,\ch{SnS} \\
\rowcolor{purple!50}
\ch{Sn2BiS2I3} 	& \num{66} 				& \num{45} 				& \num{22} 				& \num{2} 	& 2\,\ch{Sn2BiS2I3} 		$\leftrightarrow$ 2\,\ch{BiSI} 		+ 2\,\ch{SnI2} 	+ 2\,\ch{SnS} \\
\ch{Sn2BiSe2Cl3}	& \num{97} 				& \num{74} 				& \num{61} 				& \num{1} 	& 4\,\ch{Sn2BiSe2Cl3} 	$\leftrightarrow$ 2\,\ch{Bi2Se3} 	+ 6\,\ch{SnCl2} 	+ 2\,\ch{SnSe} \\
\ch{Sn2BiSe2Br3}	& \num{83} 				& \num{56} 				& \num{31} 				& \num{2} 	& 2\,\ch{Sn2BiSe2Br3} 	$\leftrightarrow$ 2\,\ch{BiSeBr} 	+ 2\,\ch{SnBr2} 	+ 2\,\ch{SnSe} \\
\ch{Sn2BiSe2I3}	& \num{92} 				& \num{55} 				& \num{37} 				& \num{2} 	& 2\,\ch{Sn2BiSe2I3} 	$\leftrightarrow$ 2\,\ch{BiSeI} 		+ 2\,\ch{SnI2} 	+ 2\,\ch{SnSe} \\
\ch{Sn2BiTe2Cl3}	& \num{136} 				& \num{123} 				& \num{80} 				& \num{3} 	& 4\,\ch{Sn2BiTe2Cl3} 	$\leftrightarrow$ 2\,\ch{SnBi2Te4}	+ 6\,\ch{SnCl2} \\
\ch{Sn2BiTe2Br3}	& \num{118} 				& \num{102}				& \num{68} 				& \num{2} 	& 2\,\ch{Sn2BiTe2Br3} 	$\leftrightarrow$ 2\,\ch{BiTeBr} 	+ 2\,\ch{SnBr2} 	+ 2\,\ch{SnTe} \\
\ch{Sn2BiTe2I3}	& \num{127} 				& \num{83} 				& \num{52} 				& \num{3} 	& 4\,\ch{Sn2BiTe2I3} 	$\leftrightarrow$ 2\,\ch{SnBi2Te4} 	+ 6\,\ch{SnI2} \\\hline
\end{tabular}}
\end{table}

\clearpage

\begin{table}[h!]
\centering
\caption*{Continuation from Table\,\ref{tab:si:hulle_decompreac}: DFT (PBE level of theory) calculated hull energies ($E_\text{hull}$), along with the corresponding decomposition reactions and reaction types, as shown in Figure\,3 of the main text, for all \protect\ch{M(II)2M(III)Ch2X3} compounds. In parentheses: Hull energies from the Materials Project\cite{materialsproject2013} if available. Colors: Purple: MMCH compounds that have been studied experimentally and/or theoretically in the literature; Cyan: lead-free MMCH compounds that are experimentally investigated in this work.}
\scalebox{.8125}{
\begin{tabular}{C{2.75cm}|C{1.275cm}C{1.275cm}C{1.275cm}||C{1.0cm}|C{9.25cm}}
~\linebreak ~\linebreak \textbf{MMCH} & \textbf{\textit{Cmcm}} \textbf{\textit{E}\textsubscript{hull}} \textbf{/\unit[per-mode=fraction]{\milli\electronvolt\per\at}} & \textbf{\textit{Cmc}2\textsubscript{1}} \textbf{\textit{E}\textsubscript{hull}} \textbf{/\unit[per-mode=fraction]{\milli\electronvolt\per\at}} & \textbf{\textit{P}2\textsubscript{1}\textit{/c}} \textbf{\textit{E}\textsubscript{hull}} \textbf{/\unit[per-mode=fraction]{\milli\electronvolt\per\at}} & ~\linebreak ~\linebreak \textbf{Type} & ~\linebreak ~\linebreak \textbf{Decomposition reaction} \\\hline\hline
\ch{Pb2InS2Cl3}	& \num{91} 				& \num{49} 				& \num{38} 				& \num{1} 	& 4\,\ch{Pb2InS2Cl3} 	$\leftrightarrow$ 2\,\ch{In2S3} 		+ 6\,\ch{PbCl2}	+ 2\,\ch{PbS} \\
\ch{Pb2InS2Br3}	& \num{103} 				& \num{45} 				& \num{31} 				& \num{1} 	& 4\,\ch{Pb2InS2Br3} 	$\leftrightarrow$ 2\,\ch{In2S3} 		+ 6\,\ch{PbBr2} 	+ 2\,\ch{PbS} \\
\ch{Pb2InS2I3} 	& \num{115} 				& \num{65} 				& \num{46} 				& \num{1} 	& 4\,\ch{Pb2InS2I3} 		$\leftrightarrow$ 2\,\ch{In2S3} 		+ 6\,\ch{PbI2} 	+ 2\,\ch{PbS} \\
\ch{Pb2InSe2Cl3}	& \num{98} 				& \num{70} 				& \num{58} 				& \num{1} 	& 4\,\ch{Pb2InSe2Cl3} 	$\leftrightarrow$ 2\,\ch{In2Se3} 	+ 6\,\ch{PbSe} 	+ 2\,\ch{PbCl2} \\
\ch{Pb2InSe2Br3}	& \num{109} 				& \num{55}				& \num{43} 				& \num{1} 	& 4\,\ch{Pb2InSe2Br3} 	$\leftrightarrow$ 2\,\ch{In2Se3} 	+ 6\,\ch{PbSe} 	+ 2\,\ch{PbBr2} \\
\ch{Pb2InSe2I3}	& \num{116} 				& \num{68} 				& \num{49} 				& \num{1} 	& 4\,\ch{Pb2InSe2I3} 	$\leftrightarrow$ 2\,\ch{PbSe} 		+ 6\,\ch{In2Se3}	+ 2\,\ch{PbI2} \\
\ch{Pb2InTe2Cl3}	& \num{60} 				& \num{84} 				& \num{68} 				& \num{4} 	& 6\,\ch{Pb2InTe2Cl3} 	$\leftrightarrow$ \ch{PbIn6Te10} 	+ 9\,\ch{PbCl2} 	+ 2\,\ch{TePb} \\
\ch{Pb2InTe2Br3}	& \num{96} 				& \num{75} 				& \num{47} 				& \num{4} 	& 6\,\ch{Pb2InTe2Br3} 	$\leftrightarrow$ \ch{PbIn6Te10} 	+ 9\,\ch{PbBr2} 	+ 2\,\ch{TePb} \\
\ch{Pb2InTe2I3}	& \num{121} 				& \num{88} 				& \num{65} 				& \num{4} 	& 6\,\ch{Pb2InTe2I3}  	$\leftrightarrow$ \ch{PbIn6Te10} 	+ 9\,\ch{PbI2} 	+ 2\,\ch{TePb} \\\hline

\ch{Pb2SbS2Cl3}	& \num{160} 				& \num{99} 				& \num{48} 				& \num{5} 	& 8\,\ch{Pb2SbS2Cl3} 	$\leftrightarrow$ \ch{Sb8(PbS5)3} 	+ 12\,\ch{PbCl2}	+ \ch{PbS} \\
\ch{Pb2SbS2Br3} 	& \num{150} 				& \num{87} 				& \num{26} 				& \num{5} 	& 8\,\ch{Pb2SbS2Br3} 	$\leftrightarrow$ \ch{Sb8(PbS5)3} 	+ 12\,\ch{PbBr2}	+ \ch{PbS} \\
\rowcolor{purple!50}
\ch{Pb2SbS2I3} 	& \num{116} 				& \num{33} (\num{48})	& \num{23} (\num{27}) 	& \num{5} 	& 8\,\ch{Pb2SbS2I3} 		$\leftrightarrow$ \ch{Sb8(PbS5)3} 	+ 12\,\ch{PbI2} 	+ \ch{PbS} \\
\ch{Pb2SbSe2Cl3}	& \num{182} 				& \num{114} 				& \num{65} 				& \num{1} 	& 4\,\ch{Pb2SbSe2Cl3} 	$\leftrightarrow$ 2\,\ch{Sb2Se3} 	+ 6\,\ch{PbCl2}	+ 2\,\ch{PbSe} \\
\ch{Pb2SbSe2Br3}	& \num{78} 				& \num{110} 				& \num{46} 				& \num{1} 	& 4\,\ch{Pb2SbSe2Br3} 	$\leftrightarrow$ 2\,\ch{Sb2Se3} 	+ 6\,\ch{PbBr2} 	+ 2\,\ch{PbSe} \\
\ch{Pb2SbSe2I3}	& \num{166} 				& \num{86} 				& \num{43} 				& \num{1} 	& 4\,\ch{Pb2SbSe2I3} 	$\leftrightarrow$ 2\,\ch{Sb2Se3} 	+ 6\,\ch{PbI2} 	+ 2\,\ch{PbSe} \\
\ch{Pb2SbTe2Cl3}	& \num{123} 				& \num{156} 				& \num{83} 				& \num{1} 	& 4\,\ch{Pb2SbTe2Cl3} 	$\leftrightarrow$ 2\,\ch{Sb2Te3} 	+ 6\,\ch{PbCl2} 	+ 2\,\ch{PbTe} \\
\ch{Pb2SbTe2Br3}	& \num{143} 				& \num{145} 				& \num{85} 				& \num{1} 	& 4\,\ch{Pb2SbTe2Br3} 	$\leftrightarrow$ 2\,\ch{Sb2Te3} 	+ 6\,\ch{PbBr2} 	+ 2\,\ch{PbTe} \\
\ch{Pb2SbTe2I3}	& \num{223} 				& \num{128} 				& \num{65} 				& \num{1} 	& 4\,\ch{Pb2SbTe2I3} 	$\leftrightarrow$ 2\,\ch{Sb2Te3} 	+ 6\,\ch{PbBr2} 	+ 2\,\ch{PbTe} \\\hline

\ch{Pb2BiS2Cl3}	& \num{125} 				& \num{67} 				& \num{53} 				& \num{3} 	& 4\,\ch{Pb2BiS2Cl3} 	$\leftrightarrow$ 2\,\ch{PbBi2S4} 	+ 6\,\ch{PbCl2} \\
\ch{Pb2BiS2Br3} 	& \num{109} 				& \num{52} 				& \num{39} 				& \num{3} 	& 4\,\ch{Pb2BiS2Br3}		$\leftrightarrow$ 2\,\ch{PbBi2S4} 	+ 6\,\ch{PbBr2} \\
\rowcolor{purple!50}
\ch{Pb2BiS2I3} 	& \num{109} 				& \num{52} 				& \num{34} 				& \num{3} 	& 4\,\ch{Pb2BiS2I3} 		$\leftrightarrow$ 2\,\ch{PbBi2S4} 	+ 6\,\ch{PbI2} \\
\ch{Pb2BiSe2Cl3}	& \num{148} 				& \num{95} 				& \num{69} 				& \num{3} 	& 4\,\ch{Pb2BiSe2Cl3} 	$\leftrightarrow$ 2\,\ch{PbBi2Se4}	+ 6\,\ch{PbCl2} \\
\ch{Pb2BiSe2Br3}	& \num{147} 				& \num{76} 				& \num{51} 				& \num{3} 	& 4\,\ch{Pb2BiSe2Br3} 	$\leftrightarrow$ 2\,\ch{PbBi2Se4} 	+ 6\,\ch{PbBr2} \\
\ch{Pb2BiSe2I3}	& \num{140} 				& \num{69} 				& \num{48} 				& \num{3} 	& 4\,\ch{Pb2BiSe2I3} 	$\leftrightarrow$ 2\,\ch{PbBi2Se4} 	+ 6\,\ch{PbI2} \\
\ch{Pb2BiTe2Cl3}	& \num{130} 				& \num{139}				& \num{81} 				& \num{3} 	& 4\,\ch{Pb2BiTe2Cl3} 	$\leftrightarrow$ 2\,\ch{PbBi2Te4} 	+ 6\,\ch{PbCl2} \\
\ch{Pb2BiTe2Br3}	& \num{182} 				& \num{121} 				& \num{7	0} 				& \num{3} 	& 4\,\ch{Pb2BiTe2Br3} 	$\leftrightarrow$ 2\,\ch{PbBi2Te4} 	+ 6\,\ch{PbBr2} \\
\ch{Pb2BiTe2I3}	& \num{183} 				& \num{99} 				& \num{92} 				& \num{3} 	& 4\,\ch{Pb2BiTe2I3} 	$\leftrightarrow$ 2\,\ch{PbBi2Te4} 	+ 6\,\ch{PbI2} \\\hline
\end{tabular}}
\end{table}

Materials Project\cite{materialsproject2013} reference data: \ch{Sn2SbS2I3} for $Cmcm$~=~mp-561134 and for $Cmc2_1$~=~mp-1219046; \ch{Sn2SbSe2I3} for $Cmc2_1$~=~mp-1219043; \ch{Pb2SbS2I3} for $Cmc2_1$~=~mp-1219558 and for $P2_1/c$~=~mp-578882.

\clearpage

\pdfbookmark[section]{Random forest regression results for predicting the hull energies in $Cmcm$, $Cmc2_1$ and $P2_1/c$}{s5}
\section{S5: Random forest regression results for predicting the hull energies in $Cmcm$, $Cmc2_1$ and $P2_1/c$}\label{sec:si:s5}

\begin{figure}[h!]
	\centering
	\includegraphics[scale=1]{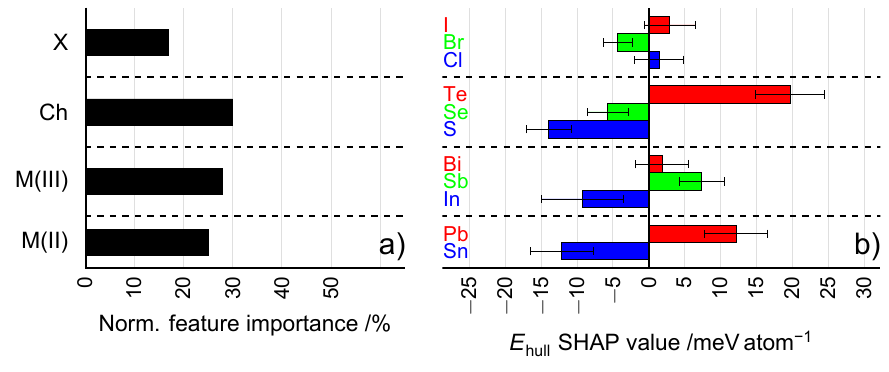}
	\caption{Summary of the feature importance for the hull energy RF model in $Cmcm$ MMCH compounds. a) Normalised atom site importance on the predicted hull energies in percent. b) Impact of elements within the \ch{M(II)}, \ch{M(III)}, \ch{Ch} and \ch{X} sites measured as mean SHAP values on the predicted base hull energy value in \unit{\milli\electronvolt\per\at}. The standard deviation for each mean SHAP value is displayed as error bars.}
	\label{fig:si:shap_cmcm}
\end{figure}

\begin{figure}[h!]
	\centering
	\includegraphics[scale=1]{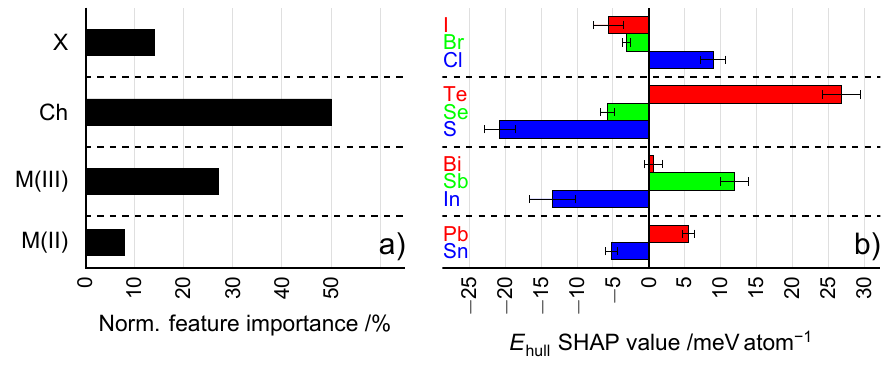}
	\caption{Summary of the feature importance for the hull energy RF model in $Cmc2_1$ MMCH compounds. a) Normalised atom site importance on the predicted hull energies in percent. b) Impact of elements within the \ch{M(II)}, \ch{M(III)}, \ch{Ch} and \ch{X} sites measured as mean SHAP values on the predicted base hull energy value in \unit{\milli\electronvolt\per\at}. The standard deviation for each mean SHAP value is displayed as error bars.}
	\label{fig:si:shap_cmc21}
\end{figure}

\begin{figure}[h!]
	\centering
	\includegraphics[scale=1]{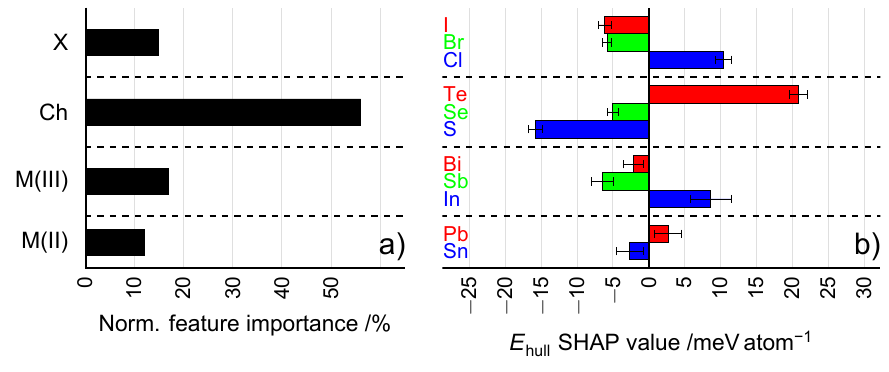}
	\caption{Summary of the feature importance for the hull energy RF model in $P2_1/c$ MMCH compounds. a) Normalised atom site importance on the predicted hull energies in percent. b) Impact of elements within the \ch{M(II)}, \ch{M(III)}, \ch{Ch} and \ch{X} sites measured as mean SHAP values on the predicted base hull energy value in \unit{\milli\electronvolt\per\at}. The standard deviation for each mean SHAP value is displayed as error bars.}
	\label{fig:si:shap_p21c}
\end{figure}

We investigated the influence of the four atomic sites (\ch{M(II)}, \ch{M(III)}, \ch{Ch}, and \ch{X}), as well as the effect of substitutional elements at these sites in MMCH compounds, using RF models and SHAP analysis, see Section\,\ref{sec:si:s1} for details. Figures\,\ref{fig:si:shap_cmcm}-\ref{fig:si:shap_p21c} a) show the feature importance derived from the RF models. Feature importance quantifies the influence of each variable within the descriptor set on the predicted values. As atomic numbers serve as compositional descriptors, the resulting importance values can be directly mapped onto the corresponding atomic sites, allowing us to evaluate their individual contributions to the hull energies. Figures\,\ref{fig:si:shap_cmcm}-\ref{fig:si:shap_p21c} b) in contrast, summarises the results of the SHAP analysis. For each RF model, a SHAP base value is first determined, corresponding to the mean predicted hull energy. The contribution of each feature to the predicted target value is then calculated relative to this base value. To enhance interpretability, we report the mean SHAP values for each element type, as the contribution of a given element depends on the specific MMCH composition, minor variations in the contributions of individual element types are observed. Table\,\ref{tab:si:design_rules_ehull} summarises the influence of individual elements at the respective atomic sites on the hull energy, as quantified by their SHAP values. A \enquote{$>$} symbol, as in \ch{Sn} $>$ \ch{Pb}, indicates that \ch{Sn} reduces the hull energy relative to the base value, whereas \ch{Pb} increases it. This means that \ch{Sn} based MMCHs lie closer to the convex hull than their \ch{Pb} counterparts and are therefore more thermodynamically stable.

\clearpage

\begin{table}[h!]
\centering
\caption{Summary of the elements’ impact on the hull energies at the atomic sites for the space groups $Cmcm$, $Cmc2_1$, $P2_1/c$, based on RF regression results with SHAP analysis, see Figures\,\ref{fig:si:shap_cmcm} ($Cmcm$), \ref{fig:si:shap_cmc21} ($Cmc2_1$) and \ref{fig:si:shap_p21c} ($P2_1/c$). }
\label{tab:si:design_rules_ehull}
\begin{tabular}{c|ccc}
				& \textbf{\textit{Cmcm}}		& \textbf{\textit{Cmc2}\textsubscript{1}} 	& \textbf{\textit{P}2\textsubscript{1}\textit{/c}} 	\\\hline\hline
\textbf{M(II)}	& Sn $>$ Pb					& Sn $>$ Pb									& Sn $>$ Pb 											\\	 
\textbf{M(III)}	& In $>$ Bi $>$ Sb			& In $>$ Bi $>$ Sb							& Sb $>$ Bi $>$ In									\\
\textbf{Ch}		& S $>$ Se $>$ Te			& S $>$ Se $>$ Te							& S $>$ Se $>$ Te									\\
\textbf{X}		& Br $>$ Cl $>$ I			& I $>$ Br $>$ Cl							& I $\geq$ Br $>$ Cl									\\
\end{tabular}
\end{table}

\clearpage

\pdfbookmark[section]{Formation energies of \protect\ch{M(II)2M(III)Ch2X3} decomposition products}{s6}
\section{S6: Formation energies of \protect\ch{M(II)2M(III)Ch2X3} decomposition products}\label{sec:si:s6}

\begin{figure}[h!]
	\centering
	\includegraphics[width=0.75\textwidth]{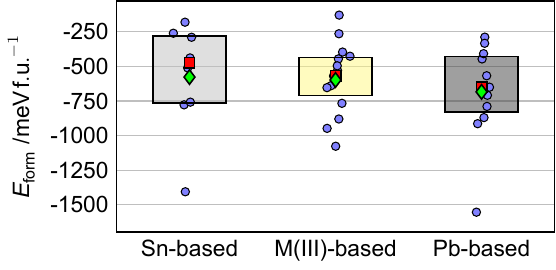}
	\caption{Boxplot for the decomposition product formation energy distribution grouped into \ch{Sn}-based, \ch{M(III)}-based, and \ch{Pb}-based compounds. The boxes represent the interquartile range with the median indicated by a red square. The green diamond represents the respective mean value. Blue dots represent the individual formation energies of the corresponding decomposition products.}
	\label{fig:si:formationenergy_decomp}
\end{figure}

\clearpage

\pdfbookmark[section]{Additional details on our broader experimental screening campaign}{s7}
\section{S7: Additional details on our broader experimental screening campaign}\label{sec:si:s7}

The thiourea (TU) based molecular ink approach reported by Nie \textit{et al.}\cite{nie2020a} was initially reproduced for the synthesis of \ch{Sn2SbS2I3} using a dimethylformamide (DMF)-based precursor system. Under these conditions, \ch{Sn2SbS2I3} was obtained with high phase purity, confirming the suitability of this route as a reference system. A detailed description of the synthesis procedure is provided in Section\,S2. Building on this protocol, analogous synthetic routes were subsequently applied to \ch{Sn2InS2Br3}, \ch{Sn2InSe2Br3}, \ch{Sn2SbS2Cl3}, \ch{Sn2SbSe2Cl3}, and \ch{Sn2InSe2Cl3}, including the substitution of TU with selenourea (SeU) for \ch{Se} based systems.

XRD analysis revealed that, for the majority of MMCH studied compounds, no quaternary phases were formed. Instead, phase formation was dominated by binary compounds, primarily \ch{SnCh} and \ch{M(III)2Ch3} phases. In one sample (see sample 7 in Table\,\ref{tab:si:mmch_exp_samples}), a ternary phase, \ch{InSeBr}, was identified. Among the investigated systems, \ch{Sn2InS2Br3} could be confirmed as a quaternary phase, whereas no quaternary phase formation was observed for the remaining MMCH compounds. A comprehensive overview of all samples, synthesis routes, annealing conditions, and identified phases are provided in Table\,\ref{tab:si:mmch_exp_samples}. To further probe the synthetic accessibility of \ch{Sn2InSe2Br3}, \ch{Sn2SbS2Cl3}, \ch{Sn2SbSe2Cl3}, and \ch{Sn2InSe2Cl3}, additional synthesis strategies were explored. In total, ten distinct approaches were evaluated across all six MMCH systems, see Section\,S2 and Table\,\ref{tab:si:mmch_exp_samples} for more details. Across all approaches, phase formation remained dominated by binary compounds, and no additional quaternary phases were identified. Incorporation of halogens into crystalline phases was generally not observed. In particular, no chlorine-containing crystalline phases were detected under any of the investigated conditions. Bromine incorporation was limited to the formation of \ch{Sn2InS2Br3} and \ch{InSeBr}, with no evidence for its incorporation into other crystalline phases. In contrast, iodine was successfully incorporated in \ch{Sn2SbS2I3}.

The influence of the annealing temperature was investigated in the range of \SIrange{150}{350}{\degreeCelsius}. Samples processed at temperatures $\leq \SI{250}{\degreeCelsius}$ frequently resulted in amorphous films, whereas higher temperatures yielded crystalline but multiphase products. For \ch{Se}-containing systems, the formation of \ch{SnSe2} was predominantly observed at annealing temperatures $\geq \SI{250}{\degreeCelsius}$. In addition, no significant differences in phase formation were observed upon variation of the substrate, including fluorine-doped tin oxide, mesoporous \ch{TiO2}, and \ch{Mo}. These observations indicate that the unsuccessful synthesis attempts are not simply due to one specific temperature or substrate condition.

\clearpage

\pdfbookmark[section]{XRD patterns and Rietfeld refinements for \protect\ch{Sn2SbS2I3}, \protect\ch{Sn2InS2Br3}, \protect\ch{Sn2InSe2Br3}, \protect\ch{Sn2SbS2Cl3}, \protect\ch{Sn2SbSe2Cl3} and \protect\ch{Sn2InSe2Cl3}}{s8}
\section{S8: XRD patterns and Rietfeld refinements for \protect\ch{Sn2SbS2I3}, \protect\ch{Sn2InS2Br3}, \protect\ch{Sn2InSe2Br3}, \protect\ch{Sn2SbS2Cl3}, \protect\ch{Sn2SbSe2Cl3} and \protect\ch{Sn2InSe2Cl3}}\label{sec:si:s8}

\clearpage

\begin{table}[h!]
\centering
\caption{Phase composition (including contributions from substrate, oxides, and reaction by-products, if present) determined based on the Rietveld refinements performed for a) \ch{Sn2SbS2I3} [see Table \ref{tab:si:rietfeld_refdata:sn2sbs2i3}], b) \ch{Sn2InS2Br3} [see Table \ref{tab:si:rietfeld_refdata:sn2ins2br3}], c) \ch{Sn2InSe2Br3} [see Table \ref{tab:si:rietfeld_refdata:sn2inse2br3}], d) \ch{Sn2SbS2Cl3} [see Table \ref{tab:si:rietfeld_refdata:sn2sbs2cl3}], e) \ch{Sn2SbSe2Cl3} [see Table \ref{tab:si:rietfeld_refdata:sn2sbse2cl3}], and f) \ch{Sn2InSe2Cl3} [see Table \ref{tab:si:rietfeld_refdata:sn2inse2cl3}].}
\label{tab:si:phase_comp}
\scalebox{.8125}{
\begin{tabular}{c|C{1.6cm}C{1.65cm}C{2.0cm}C{1.85cm}C{1.55cm}C{2.3cm}|C{1.7cm}C{0.7cm}C{0.9cm}}
 & \textbf{MMCH} \textbf{/\unit{\percent}} & \textbf{M(II)Ch} \textbf{/\unit{\percent}} & \textbf{M(II)\textsubscript{2}Ch\textsubscript{3}} \textbf{/\unit{\percent}} & \textbf{M(II)Ch\textsubscript{2}} \textbf{/\unit{\percent}} & \textbf{M(II)X\textsubscript{2}} \textbf{/\unit{\percent}} & \textbf{M(III)\textsubscript{2}Ch\textsubscript{3}} \textbf{/\unit{\percent}} & \textbf{M(II)O\textsubscript{2}} \textbf{/\unit{\percent}} & \textbf{Mo} \textbf{/\unit{\percent}} & \textbf{InCl} \textbf{/\unit{\percent}}\\\hline\hline
\rowcolor{green!30}
\ch{Sn2SbS2I3} 	& \num{84.3} 	& \num{12.2} 	& -/- 			& -/- 			& -/- 			& \num{1.0} 		& \num{2.5} 		& -/- 			& -/-\\
\rowcolor{green!30}
\ch{Sn2InS2Br3} 	& \num{19.2} 	& \num{44.2} 	& -/- 			& -/- 			& \num{14.5} 	& \num{16.6} 	& -/- 			& -/- 			& \num{5.5} \\
\ch{Sn2InSe2Br3}	& -/- 			& \num{2.1} 		& -/- 			& \num{21.1} 	& -/- 			& \num{0.9} 		& -/-			& \num{75.9}		& -/- \\
\ch{Sn2SbS2Cl3}	& \num{1.3} 		& \num{18.5} 	& \num{71.4}		& -/- 			& \num{2.9} 		& \num{5.9} 		& -/- 			& -/- 			& -/- \\
\ch{Sn2SbSe2Cl3}	& \num{0.3} 		& \num{95.3} 	& -/- 			& -/- 			& -/- 			& \num{4.4} 		& -/- 			& -/- 			& -/- \\
\ch{Sn2InSe2Cl3}	& \num{0.4} 		& \num{61.5} 	& -/- 			& -/- 			& -/- 			& \num{8.4} 		& -/-		   	& \num{29.7}		& -/- \\
\end{tabular}}
\end{table}

\begin{table}[h!]
\centering
\caption{Crystallographic data and Rietveld refinement parameters for a \protect\ch{Sn2SbS2I3} film prepared from a \ch{SnI2} + \num{0.7}\,\ch{SbCl3} + \num{1.8}\,TU precursor solution in DMF on a FTO/\ch{TiO2}/mp-\ch{TiO2} substrate and annealed for \SI{5}{\min} at \SI{300}{\celsius}.}
\label{tab:si:rietfeld_refdata:sn2sbs2i3}
\scalebox{.8125}{
\begin{tabular}{C{3.0cm}|C{3.1cm}|C{2.25cm}C{2.25cm}C{2.25cm}C{2.25cm}C{2.25cm}C{2.25cm}}
 & \textbf{MMCH phase:} \textbf{Sn\textsubscript{2}SbS\textsubscript{2}I\textsubscript{3}} & \textbf{Phase 2} \textbf{SnS} & \textbf{Phase 3} \textbf{SnI\textsubscript{2}} & \textbf{Phase 4} \textbf{Sb\textsubscript{2}S\textsubscript{3}} & \textbf{Phase 5} \textbf{SnO\textsubscript{2}} & \textbf{Phase 6} \linebreak -/-\\\hline\hline
\textbf{Space group}		 				& $Cmcm$ 		& $Pnma$			& $C2/m$ 		& $Pnma$			& $P4_2/mnm$ 		& \\
\textbf{\textit{a}} /\unit{\angstrom} 	& \num{4.26} 	& \num{11.29} 	& \num{14.78}	& \num{11.80}	& \num{4.75}			& \\
\textbf{\textit{b}} /\unit{\angstrom} 	& \num{13.99} 	& \num{3.95} 	& \num{4.52}		& \num{3.84} 	& \num{4.75}			& \\
\textbf{\textit{c}} /\unit{\angstrom} 	& \num{16.34} 	& \num{4.60}		& \num{10.98}	& \num{11.36}	& \num{3.20}			& \\
$\boldsymbol{\alpha}$ /\unit{\degree} 	& \num{90.0} 	& \num{90.0}		& \num{90.0}		& \num{90.0} 	& \num{90.0}			& \\
$\boldsymbol{\beta}$ /\unit{\degree} 	& \num{90.0} 	& \num{90.0}		& \num{91.8}		& \num{90.0} 	& \num{90.0}			& \\
$\boldsymbol{\gamma}$ /\unit{\degree}	& \num{90.0} 	& \num{90.0}		& \num{90.0}		& \num{90.0} 	& \num{90.0}			& \\
\textbf{Volume} /\unit{\angstrom\cubed} 	& \num{974.5}	& \num{204.7}	& \num{733.1}	& \num{516.1}	& \num{72.4}			& \\
$\boldsymbol{\rho}$ /\unit{\gram\per\cm\cubed} %
										& \num{5.541} 	& \num{7.179} 	& \num{5.192} 	& \num{2.185} 	& \num{1.342} 		& \\
\textbf{\textit{R}\textsubscript{B}} /\unit{\percent} %
										& \num{14.8} 	& \num{26.5} 	& \num{49.5} 	& \num{45.2} 	& \num{30.2} 		& \\\hline\hline
\textbf{Conv. \textit{R}\textsubscript{p}} /\unit{\percent} 	& \multicolumn{6}{c}{\num{36.8}} \\
\textbf{Conv. \textit{R}\textsubscript{wp}} /\unit{\percent} 	& \multicolumn{6}{c}{\num{38.1}} \\
\textbf{Conv. \textit{R}\textsubscript{e}} /\unit{\percent} 	& \multicolumn{6}{c}{\num{11.4}} \\
$\boldsymbol{\chi^2}$				 						& \multicolumn{6}{c}{\num{11.3}} \\
\end{tabular}}
\end{table}

\begin{table}[h!]
\centering
\caption{Crystallographic data and Rietveld refinement parameters for a \protect\ch{Sn2InS2Br3} film prepared from a \ch{SnBr2} + \num{0.7}\,\ch{InCl3} + \num{1.8}\,TU precursor solution in DMF on a FTO substrate and annealed for \SI{5}{\min} at \SI{300}{\celsius}.}
\label{tab:si:rietfeld_refdata:sn2ins2br3}
\scalebox{.8125}{
\begin{tabular}{C{3.0cm}|C{3.1cm}|C{2.25cm}C{2.25cm}C{2.25cm}C{2.25cm}C{2.25cm}C{2.25cm}}
 & \textbf{MMCH phase:} \textbf{Sn\textsubscript{2}InS\textsubscript{2}Br\textsubscript{3}} & \textbf{Phase 2} \textbf{SnS} & \textbf{Phase 3} \textbf{SnBr\textsubscript{2}} & \textbf{Phase 4} \textbf{In\textsubscript{2}S\textsubscript{3}} & \textbf{Phase 5} \textbf{InCl} & \textbf{Phase 6} \linebreak -/-\\\hline\hline
\textbf{Space group}		 				& $Cmcm$ 		& $Pnma$			& $Pnma$ 		& $R\overline{3}c$	& $P2_13$ 			& \\
\textbf{\textit{a}} /\unit{\angstrom} 	& \num{4.33} 	& \num{11.19} 	& \num{8.70}		& \num{6.57}			& \num{12.42}		& \\
\textbf{\textit{b}} /\unit{\angstrom} 	& \num{14.29} 	& \num{3.87} 	& \num{4.23}		& \num{6.57} 		& \num{12.42}		& \\
\textbf{\textit{c}} /\unit{\angstrom} 	& \num{14.64} 	& \num{4.51}		& \num{10.80}	& \num{18.10}		& \num{12.42}		& \\
$\boldsymbol{\alpha}$ /\unit{\degree} 	& \num{90.0} 	& \num{90.0}		& \num{90.0}		& \num{90.0} 		& \num{90.0}			& \\
$\boldsymbol{\beta}$ /\unit{\degree} 	& \num{90.0} 	& \num{90.0}		& \num{90.0}		& \num{90.0} 		& \num{90.0}			& \\
$\boldsymbol{\gamma}$ /\unit{\degree} 	& \num{90.0} 	& \num{90.0}		& \num{90.0}		& \num{120.0} 		& \num{90.0}			& \\
\textbf{Volume} /\unit{\angstrom\cubed} 	& \num{906.7}	& \num{195.0}	& \num{397.9}	& \num{677.1}		& \num{1917.8}		& \\
$\boldsymbol{\rho}$ /\unit{\gram\per\cm\cubed} %
										& \num{7.509} 	& \num{6.571} 	& \num{5.666} 	& \num{4.048} 		& \num{3.052} 		& \\
\textbf{\textit{R}\textsubscript{B}} /\unit{\percent} %
										& \num{13.4} 	& \num{8.2} 		& \num{21.2} 	& \num{9.9}	 		& \num{16.0} 		& \\\hline\hline
\textbf{Conv. \textit{R}\textsubscript{p}} /\unit{\percent} 	& \multicolumn{6}{c}{\num{34.7}} \\
\textbf{Conv. \textit{R}\textsubscript{wp}} /\unit{\percent} 	& \multicolumn{6}{c}{\num{36.0}} \\
\textbf{Conv. \textit{R}\textsubscript{e}} /\unit{\percent} 	& \multicolumn{6}{c}{\num{16.4}} \\
$\boldsymbol{\chi^2}	$				 						& \multicolumn{6}{c}{\num{4.9}} \\
\end{tabular}}
\end{table}

\begin{table}[h!]
\centering
\caption{Crystallographic data and Rietveld refinement parameters for a \protect\ch{Sn2InSe2Br3} film prepared from a \ch{SnBr2} + \num{0.7}\,\ch{InCl3} + \ch{1.8}\,SeU precursor solution in DMF on a \ch{Mo} substrate and annealed for \SI{5}{\min} at \SI{350}{\celsius}.}
\label{tab:si:rietfeld_refdata:sn2inse2br3}
\scalebox{.8125}{
\begin{tabular}{C{3.0cm}|C{3.1cm}|C{2.25cm}C{2.25cm}C{2.25cm}C{2.25cm}C{2.25cm}C{2.25cm}}
 & \textbf{MMCH phase:} \textbf{Sn\textsubscript{2}InSe\textsubscript{2}Br\textsubscript{3}} & \textbf{Phase 2:} \textbf{SnSe} & \textbf{Phase 3:} \textbf{SnSe\textsubscript{2}} & \textbf{Phase 4:}   \textbf{SnBr\textsubscript{2}} & \textbf{Phase 5:} \textbf{In\textsubscript{2}Se\textsubscript{3}} & \textbf{Phase 6:} \textbf{Mo}\\\hline\hline
\textbf{Space group}		 				& $Cmcm$ 		& $Pnma$			& $P\overline{3}m1$ 	& $Pnmac$			& $R\overline{3}m$	& $Im\overline{3}m$ \\
\textbf{\textit{a}} /\unit{\angstrom} 	& \num{4.34} 	& \num{11.37} 	& \num{3.79}			& \num{8.76}			& \num{4.05}			& \num{3.14} \\
\textbf{\textit{b}} /\unit{\angstrom} 	& \num{14.79} 	& \num{4.18} 	& \num{3.79}			& \num{4.24} 		& \num{4.05}			& \num{3.14} \\
\textbf{\textit{c}} /\unit{\angstrom} 	& \num{15.05} 	& \num{4.44}		& \num{6.24}			& \num{10.170}		& \num{29.42}		& \num{3.14} \\
$\boldsymbol{\alpha}$ /\unit{\degree} 	& \num{90.0} 	& \num{90.0}		& \num{90.0}			& \num{90.0} 		& \num{90.0}			& \num{90.0} \\
$\boldsymbol{\beta}$ /\unit{\degree} 	& \num{90.0} 	& \num{90.0}		& \num{90.0}			& \num{90.0} 		& \num{90.0}			& \num{90.0} \\
$\boldsymbol{\gamma}$ /\unit{\degree}	& \num{90.0} 	& \num{90.0}		& \num{120.0}		& \num{90.0} 		& \num{120.0}		& \num{90.0} \\
\textbf{Volume} /\unit{\angstrom\cubed} 	& \num{965.3}	& \num{211.1}	& \num{77.6}			& \num{397.6}		& \num{418.4}		& \num{31.1} \\
$\boldsymbol{\rho}$ /\unit{\gram\per\cm\cubed} %
										& \num{44.672} 	& \num{4.743} 	& \num{5.573} 		& \num{4.652} 		& \num{3.040} 		& \num{10.249} \\
\textbf{\textit{R}\textsubscript{B}} /\unit{\percent} %
										& \num{77.0} 	& \num{32.9} 	& \num{21.9} 		& \num{98.2}	 		& \num{64.0} 		& \num{15.2} \\\hline\hline
\textbf{Conv. \textit{R}\textsubscript{p}} /\unit{\percent} 	& \multicolumn{6}{c}{\num{35.6}} \\
\textbf{Conv. \textit{R}\textsubscript{wp}} /\unit{\percent} 	& \multicolumn{6}{c}{\num{34.3}} \\
\textbf{Conv. \textit{R}\textsubscript{e}} /\unit{\percent} 	& \multicolumn{6}{c}{\num{16.8}} \\
$\boldsymbol{\chi^2}	$				 						& \multicolumn{6}{c}{\num{4.2}} \\
\end{tabular}}
\end{table}

\begin{table}[h!]
\centering
\caption{Crystallographic data and Rietveld refinement parameters for a \protect\ch{Sn2SbS2Cl3} film prepared from a \ch{SnCl2} + \num{0.7}\,\ch{SbCl3} + \num{1.8}\,TU precursor solution in DMF on a FTO substrate and annealed for \SI{5}{\min} at \SI{300}{\celsius}.}
\label{tab:si:rietfeld_refdata:sn2sbs2cl3}
\scalebox{.8125}{
\begin{tabular}{C{3.0cm}|C{3.1cm}|C{2.25cm}C{2.25cm}C{2.25cm}C{2.25cm}C{2.25cm}C{2.25cm}}
 & \textbf{MMCH phase:} \textbf{Sn\textsubscript{2}SbS\textsubscript{2}Cl\textsubscript{3}} & \textbf{Phase 2:} \textbf{SnS} & \textbf{Phase 3:} \textbf{Sn\textsubscript{2}S\textsubscript{3}} & \textbf{Phase 4:} \textbf{SnCl\textsubscript{2}} & \textbf{Phase 5:} \textbf{Sb\textsubscript{2}S\textsubscript{3}} & \textbf{Phase 6:} \linebreak -/-\\\hline\hline
\textbf{Space group}		 				& $Cmcm$ 		& $Pnma$			& $Pnma$ 		& $Pnma$				& $Pnma$ 			& \\
\textbf{\textit{a}} /\unit{\angstrom} 	& \num{4.35} 	& \num{12.29} 	& \num{8.98}		& \num{7.79}			& \num{11.17}		& \\
\textbf{\textit{b}} /\unit{\angstrom} 	& \num{14.30} 	& \num{3.91} 	& \num{3.78}		& \num{9.23} 		& \num{3.78}			& \\
\textbf{\textit{c}} /\unit{\angstrom} 	& \num{16.17} 	& \num{4.36}		& \num{14.64}	& \num{4.44}			& \num{12.65}		& \\
$\boldsymbol{\alpha}$ /\unit{\degree} 	& \num{90.0} 	& \num{90.0}		& \num{90.0}		& \num{90.0} 		& \num{90.0}			& \\
$\boldsymbol{\beta}$ /\unit{\degree} 	& \num{90.0} 	& \num{90.0}		& \num{90.0}		& \num{90.0} 		& \num{90.0}			& \\
$\boldsymbol{\gamma}$ /\unit{\degree}	& \num{90.0} 	& \num{90.0}		& \num{90.0}		& \num{90.0} 		& \num{90.0}			& \\
\textbf{Volume} /\unit{\angstrom\cubed} 	& \num{1004.7}	& \num{209.9}	& \num{496.8}	& \num{318.8}		& \num{534.2}		& \\
$\boldsymbol{\rho}$ /\unit{\gram\per\cm\cubed} %
										& \num{2.693} 	& \num{4.052} 	& \num{7.268} 	& \num{1.479} 		& \num{2.789} 		& \\
\textbf{\textit{R}\textsubscript{B}} /\unit{\percent} %
										& \num{26.5} 	& \num{18.1}		& \num{12.6} 	& \num{41.2}	 		& \num{27.1} 		& \\\hline\hline
\textbf{Conv. \textit{R}\textsubscript{p}} /\unit{\percent} 	& \multicolumn{6}{c}{\num{37.2}} \\
\textbf{Conv. \textit{R}\textsubscript{wp}} /\unit{\percent} 	& \multicolumn{6}{c}{\num{35.5}} \\
\textbf{Conv. \textit{R}\textsubscript{e}} /\unit{\percent} 	& \multicolumn{6}{c}{\num{19.3}} \\
$\boldsymbol{\chi^2}	$				 						& \multicolumn{6}{c}{\num{3.4}} \\
\end{tabular}}
\end{table}

\begin{table}[h!]
\centering
\caption{Crystallographic data and Rietveld refinement parameters for a \protect\ch{Sn2SbSe2Cl3} film prepared from a \num{1.2}\,\ch{SnCl2} + \num{0.7}\,\ch{SbCl3} + \num{1.8}\,SeU precursor solution in DMF on a FTO substrate and annealed for \SI{5}{\min} at \SI{300}{\celsius}.}
\label{tab:si:rietfeld_refdata:sn2sbse2cl3}
\scalebox{.8125}{
\begin{tabular}{C{3.0cm}|C{3.1cm}|C{2.25cm}C{2.25cm}C{2.25cm}C{2.25cm}C{2.25cm}C{2.25cm}}
 & \textbf{MMCH phase:} \textbf{Sn\textsubscript{2}SbSe\textsubscript{2}Cl\textsubscript{3}} & \textbf{Phase 2:} \textbf{SnSe} & \textbf{Phase 3:} \textbf{SnCl\textsubscript{2}} & \textbf{Phase 4:} \textbf{Sb\textsubscript{2}Se\textsubscript{3}} & \textbf{Phase 5:} \linebreak -/- & \textbf{Phase 6:} \linebreak -/-\\\hline\hline
\textbf{Space group}		 				& $Cmcm$ 		& $Pnma$			& $Pnma$ 		& $Pnma$				&  & \\
\textbf{\textit{a}} /\unit{\angstrom} 	& \num{4.15} 	& \num{11.42} 	& \num{7.65}		& \num{11.79}		&  & \\
\textbf{\textit{b}} /\unit{\angstrom} 	& \num{12.98} 	& \num{4.14} 	& \num{9.36}		& \num{3.87} 		&  & \\
\textbf{\textit{c}} /\unit{\angstrom} 	& \num{14.89} 	& \num{4.28}		& \num{4.49}		& \num{11.32}		&  & \\
$\boldsymbol{\alpha}$ /\unit{\degree} 	& \num{90.0} 	& \num{90.0}		& \num{90.0}		& \num{90.0} 		&  & \\
$\boldsymbol{\beta}$ /\unit{\degree} 	& \num{90.0} 	& \num{90.0}		& \num{90.0}		& \num{90.0} 		&  & \\
$\boldsymbol{\gamma}$ /\unit{\degree}	& \num{90.0} 	& \num{90.0}		& \num{90.0}		& \num{90.0} 		&  & \\
\textbf{Volume} /\unit{\angstrom\cubed} 	& \num{802.2}	& \num{202.6}	& \num{321.5}	& \num{517.5}		&  & \\
$\boldsymbol{\rho}$ /\unit{\gram\per\cm\cubed} %
										& \num{8.362} 	& \num{5.271} 	& \num{1.486} 	& \num{8.249} 		&  & \\
\textbf{\textit{R}\textsubscript{B}} /\unit{\percent} %
										& \num{52.5} 	& \num{12.2}		& \num{44.7} 	& \num{55.5}	 		&  & \\\hline\hline
\textbf{Conv. \textit{R}\textsubscript{p}} /\unit{\percent} 	& \multicolumn{6}{c}{\num{50.8}} \\
\textbf{Conv. \textit{R}\textsubscript{wp}} /\unit{\percent} 	& \multicolumn{6}{c}{\num{43.9}} \\
\textbf{Conv. \textit{R}\textsubscript{e}} /\unit{\percent} 	& \multicolumn{6}{c}{\num{25.7}} \\
$\boldsymbol{\chi^2}	$				 						& \multicolumn{6}{c}{\num{2.9}} \\
\end{tabular}}
\end{table}

\clearpage

\begin{table}[h!]
\centering
\caption{Crystallographic data and Rietveld refinement parameters for a \protect\ch{Sn2InSe2Cl3} film prepared from a \ch{SnCl2} + \num{0.7}\,\ch{InCl3} + \num{1.8}\,SeU precursor solution in DMF on a \ch{Mo} substrate and annealed for \SI{5}{\min} at \SI{350}{\celsius}.}
\label{tab:si:rietfeld_refdata:sn2inse2cl3}
\scalebox{.8125}{
\begin{tabular}{C{3.0cm}|C{3.1cm}|C{2.25cm}C{2.25cm}C{2.25cm}C{2.25cm}C{2.25cm}C{2.25cm}}
 & \textbf{MMCH phase:} \textbf{Sn\textsubscript{2}InSe\textsubscript{2}Cl\textsubscript{3}} & \textbf{Phase 2:} \textbf{SnSe} & \textbf{Phase 3:} \textbf{SnCl\textsubscript{2}} & \textbf{Phase 4:} \textbf{In\textsubscript{2}Se\textsubscript{3}} & \textbf{Phase 5:} \textbf{Mo} & \textbf{Phase 6:} \linebreak -/-\\\hline\hline
\textbf{Space group}		 				& $Cmcm$ 		& $Pnma$			& $Pnma$ 		& $R\overline{3}m$		& $Im\overline{3}m$ 		& \\
\textbf{\textit{a}} /\unit{\angstrom} 	& \num{4.39} 	& \num{11.54} 	& \num{7.79}		& \num{4.05}				& \num{3.14} 			& \\
\textbf{\textit{b}} /\unit{\angstrom} 	& \num{13.48} 	& \num{4.22} 	& \num{9.21}		& \num{4.05} 			& \num{3.14} 			& \\
\textbf{\textit{c}} /\unit{\angstrom} 	& \num{15.46} 	& \num{4.35}		& \num{4.43}		& \num{29.41}			& \num{3.14} 			& \\
$\boldsymbol{\alpha}$ /\unit{\degree} 	& \num{90.0} 	& \num{90.0}		& \num{90.0}		& \num{90.0} 			& \num{90.0} 			& \\
$\boldsymbol{\beta}$ /\unit{\degree} 	& \num{90.0} 	& \num{90.0}		& \num{90.0}		& \num{90.0} 			& \num{90.0} 			& \\
$\boldsymbol{\gamma}$ /\unit{\degree}	& \num{90.0} 	& \num{90.0}		& \num{90.0}		& \num{120.0} 			& \num{90.0} 			& \\
\textbf{Volume} /\unit{\angstrom\cubed} 	& \num{915.9}	& \num{211.8}	& \num{317.9}	& \num{417.8}			& \num{31.2} 			& \\
$\boldsymbol{\rho}$ /\unit{\gram\per\cm\cubed} %
										& \num{8.271} 	& \num{4.951} 	& \num{3.962} 	& \num{3.266} 			& \num{10.223}			& \\
\textbf{\textit{R}\textsubscript{B}} /\unit{\percent} %
										& \num{59.4} 	& \num{16.2}		& \num{10.3} 	& \num{45.7}	 			& \num{11.2}				& \\\hline\hline
\textbf{Conv. \textit{R}\textsubscript{p}} /\unit{\percent} 	& \multicolumn{6}{c}{\num{40.3}} \\
\textbf{Conv. \textit{R}\textsubscript{wp}} /\unit{\percent} 	& \multicolumn{6}{c}{\num{32.6}} \\
\textbf{Conv. \textit{R}\textsubscript{e}} /\unit{\percent} 	& \multicolumn{6}{c}{\num{22.3}} \\
$\boldsymbol{\chi^2}	$				 						& \multicolumn{6}{c}{\num{2.1}} \\
\end{tabular}}
\end{table}

\clearpage

\pdfbookmark[section]{References}{references}
\begingroup
\small
\bibliography{lib_manuscript}
\endgroup